\documentclass[12pt,a4paper]{article}
\usepackage[a4paper, left=1.5cm, top=3.8cm, right=1.5cm, bottom=3.25cm, marginparsep=5pt, marginparwidth=20mm, headheight=22.53pt]{geometry}

\usepackage{newtxtext,newtxmath}
\usepackage{comment}
\usepackage{makecell}

\usepackage[T1]{fontenc}
\usepackage{rotating}

\DeclareRobustCommand{\VAN}[3]{#2}
\let\VANthebibliography\thebibliography
\def\thebibliography{\DeclareRobustCommand{\VAN}[3]{##3}\VANthebibliography}

\usepackage{graphicx} 
\usepackage{amsmath}
\usepackage{algorithm,algcompatible}
\algnewcommand\INPUT{\item[\textbf{Input:}]}
\algnewcommand\OUTPUT{\item[\textbf{Output:}]}
\usepackage{natbib}
\usepackage{url}
\usepackage{hyperref}
\usepackage{adjustbox}
\usepackage{authblk}

\title{A novel conjunction filter based on the minimum distance
  between perturbed trajectories}

\author[1]{A.S. Rivero}
\author[2]{G. Ba\`u}
\author[3]{R. Vazquez}
\author[4]{C. Bombardelli}
\affil[1,3]{Department of Aerospace Engineering, Universidad de Sevilla}
\affil[2]{Department of Mathematics, Universit\`a di Pisa}
\affil[4]{Department of Applied Physics, Universidad Polit\'ecnica de Madrid}

\date{}

\begin{document}
\maketitle

\begin{abstract}
The increasing congestion in the near-Earth space environment has
amplified the need for robust and efficient conjunction analysis
techniques including the computation of the minimum distance between
orbital paths in the presence of perturbations.  After showing that
classical Minimum Orbit Intersection Distance (MOID) computation
schemes are unsuitable to treat Earth orbiting objects, the article
presents an analytical approach to provide a more accurate estimate of
the true distance between perturbed trajectories by incorporating the
effect of zonal harmonics of arbitrary order. Cook's linear secular
theory for the motion of the eccentricity vector is extended to
include higher order eccentricity effects and applied to the
computation of the minimum and maximum radii attained by two orbits at
their mutual nodes, which can be employed to estimate the true
distance between the two orbital paths and to establish an efficient
algorithm for determining or excluding potential
conjunctions. Extensive testing and validation are conducted using a
high-fidelity propagator and a comprehensive dataset of resident space
objects. The results demonstrate an accuracy below the km level for
the orbit distance computation in 99\% of cases, which enables
high-efficiency conjunction filtering.
\end{abstract}


\section{Introduction}

The near-Earth space environment is undergoing a profound
transformation in the ``new space era'' \citep{Muelhaupt19}
characterized by a significant surge in space traffic, the deployment
of extensive satellite constellations, and the escalating
proliferation of space debris. This evolution presents increasingly
formidable challenges for space management, particularly concerning
the safety and sustainability of space operations. Conjunction
Analysis (CA) emerges as a pivotal tool in this context, offering
methodologies and processes to evaluate and mitigate collision risks
in orbit. By combining advanced detection techniques, risk assessment,
and collision avoidance decision-making, CA plays a crucial role in
safeguarding space assets and ensuring the long-term sustainability of
the space environment \citep{Kerr21}.

At present, operators, space agencies, and providers of Space
Situational Awareness (SSA) and Space Surveillance and Tracking (SST)
services are primarily focused on addressing what is known as the
\emph{one-vs-all} conjunction screening problem \citep{Kerr21}. This
entails a meticulous scrutiny of potential conjunctions between one
single active satellite and the extensive catalog of space
objects. While this empowers operators to make well-informed decisions
regarding Collision Avoidance Maneuvers (CAM) to protect their assets,
it falls short in addressing the myriad of unexplored collisions among
the rest of the catalogued RSO population. These collisions pose a
significant threat, potentially leading to the formation of hazardous
fragment clouds capable of causing substantial damage or even the
complete loss of operational satellites. Achieving a comprehensive
understanding of the near-Earth space environment necessitates
considering conjunctions among all conceivable combinations of
cataloged objects, commonly referred to as the \emph{all-vs-all}
scenario \citep{Stevenson2023}. However, implementing this broader
perspective presents formidable computational challenges due to the
ever-increasing number of potential conjunction pairs.

Previous research endeavors have sought solutions to these challenges
through the exploration of diverse filtering processes \citep{Wood,
  Escobar, Casanova, SO_filter} and the utilization of parallel
computing techniques \citep{Healy}. Typically, a sequence of three
filters is applied in cascade starting from an apogee-perigee filter
\citep[recently improved taking into account the effect of zonal
  harmonics in][]{SO_filter} followed by an orbit path filter and a
time filter \citep[see][]{Hoots84,Casanova}. An alternative approach
in which the apogee-perigee filter is followed by a different set of
``sieves'' has also been proposed \citep{smartsieve}. The application
of filters to eliminate pairs of Resident Space Objects (RSOs) with no
collision risk has emerged as one of the most widely adopted
techniques for expediting computational processes in conjunction
analyses.

The widely adopted \emph{orbit path filter}, which is the main focus
of the present work, hinges on the distance between two trajectories
irrespective of the actual position of the two objects along them.  A
pair of objects whose orbits have a mutual distance above a safe and
carefully established threshold have a negligible probability of
collision and can thus be eliminated (i.e., filtered out) from further
computation.

For Keplerian orbits, the function of the distance between two orbits
has been thoroughly investigated in the literature
\citep{Gronchi02,Gronchi05,hedo2018minimum,hedo2020minimum,gronchi2023revisiting}. The
absolute minimum of that function, known as the Minimum Orbit
Intersection Distance or MOID, is of great importance in astronomy and
widely employed to assess the collision risk between astronomical
bodies. In principle, a Keplerian MOID computation scheme can be
applied to perturbed orbits by considering osculating orbital
elements. However, as it will be shown in the present work, the
corresponding ``osculating'' MOID function falls short from providing
a good estimation of the true distance (i.e., the ``true'' MOID)
between RSO pairs in low-Earth orbit. This aspect was already noticed
by the works of \citet{Hoots84} and \citet{Wood} where the
applicability of purely Keplerian modeling for the implementation of
path filters was shown to be problematic.

In order to address these challenges and increase the robustness of
the orbit path filter, different approaches have been proposed in the
literature. Numerical iteration and sampling taking into account
$J_2$-induced orbital plane precession have been proposed by
\cite{Hoots84} and \cite{Wood} in order to adapt an orbit path filter
to the effect of environmental perturbations. In spite of these
efforts, \citet{Wood} conclude by not recommending the use of a path
filter, which is is labelled as ``troublesome''. \citet{Alfano12}
introduces a geometrical approach, offering versatility by enabling
users to specify different in-plane and out-of-plane bounds for the
path filter. However, in order to fully eliminate false negatives,
relatively large buffer distances (above 30 km) are incorporated.

The approach proposed in the present work, leading to the design of a
``space occupancy path conjunction filter'' (SOP-filter), is based on
three fundamental ideas. The first is to avoid treating RSO pairs with
mutual inclination outside of the range $[10,170]$ degrees and orbit
eccentricities exceeding $0.1$. The former condition ensures that the
MOID falls in the vicinity of the line of intersection between the two
orbit planes, the latter allows us an analytical treatment of the
problem.  As shown in the results section, and taking the current
space catalogue, pairs not satisfying these conditions (and that will
clearly remain unfiltered) amount to about $5\%$ of total number of
pairs analysed by the new filter. The second idea is to estimate for
each object of a given pair the orbit radius at the mutual nodes using
an efficient analytical approach based on the extension of the work by
\citet{SO_filter}, which incorporates a complete zonal gravitational
perturbation model and considers this effect over a selected
conjunction screening timeframe. The third idea is to estimate for
each of the two mutual nodes the minimum and maximum values taken by
the orbit radius across a typical 5-days conjunction, thus defining
for each object two radial intervals, and check whether the intervals
relative to the same pair of mutual nodes overlap. If no overlap is
detected the pair can be filtered out.

The article's structure is as follows. Section~\ref{sec:true_MOID}
introduces the definition of true MOID between perturbed orbit paths
and points out the inadequacy of a Keplerian MOID function to estimate
this quantity. Section~\ref{sec:radius} is devoted to the computation
of an analytical expression for the evolution of the orbit radius
under the effect of a complete zonal perturbation model. It provides a
generalization of a similar expression obtained by \cite{SO_filter} by
including higher order eccentricity effects that are crucial for the
correct implementation of the newly proposed filter and the influence
of all even zonal harmonics by following the theory of
\citet{cook1992}. Subsequently, Section~\ref{sec:SOP_filter}
delineates the development and analytical implementation of a new
orbit path filter based on the short-term SO model. Then,
Section~\ref{sec:Results} offers an analysis of the filter's
performance, focusing on reliability and effectiveness. These results
are corroborated using a high-fidelity propagator. The article
concludes with some final remarks in Section~\ref{sec:conclusions}.

\section{True vs. Keplerian osculating MOID}
\label{sec:true_MOID}

The implementation of an effective orbit path filter relies on the
concept of minimum orbit intersection distance (MOID) between two,
generally non-Keplerian trajectories $T_{A}$ and $T_{B}$. Rigorously,
that distance is the minimum distance bewteen any two of their
respective points $P_{A}$ and $P_{B}$ and will be referred here as
true MOID:
\[
\mathrm{true\,MOID}=\min_{P_{A}\in T_{A},P_{B}\in T_{B}}\left\Vert {P}_{A}-{P}_{B}\right\Vert.
\]
The preceding quantity can be estimated by considering all possible
pairs of points on the numerically propagated trajectories and
selecting the pair with minimum distance. If the two trajectories are
sampled with sufficient spatial resolution that estimation can be
taken as an accurate value for their true MOID.

In order to save computational time, one might want to consider
computing the MOID of the two osculating Keplerian trajectories at
different instants of time by employing their osculating orbital
elements and exploiting high efficiency MOID computation tools
available in the literature \citep[e.g.,][]{gronchi2023revisiting}. In
doing this, one ends up with a Keplerian osculating MOID as a function
of time and can consider the minimum of that function as a proxy for
the true MOID \citep[see for instance][]{Alfano12, Casanova,
  bonaccorsi2023software}. Unfortunately, as it will be shown in the
following examples, the minimum of the Keplerian osculating MOID is
generally not a good estimate for the true MOID.

\begin{table}
  \centering 
  \begin{tabular}{lrrr} 
    \hline 
    RSO (NORAD) & 41732 & 42775 & 41460\\ \hline
    $a$ (km) & 6,839.44 & 6,867.10 & 6,875.36\\
    $e$ & 0.0023 & 0.0019 & 0.0125 \\
    $i$ (deg) & 97.36 & 97.20 & 98.27\\
    $\omega$ (deg) & 120.14 & 139.31  & 129.61 \\
    $\Omega$ (deg) & 229.73 & 339.11 & 129.53\\ 
    \hline 
  \end{tabular}
  \caption{Osculating orbital elements at epoch JD 2,459,885.88 for
    the RSOs considered in Fig.~\ref{fig:MOID37}.}
  \label{tab:NORAD_data}
\end{table}

\begin{figure}[h]
  \centering
  \includegraphics[width=0.49\textwidth]{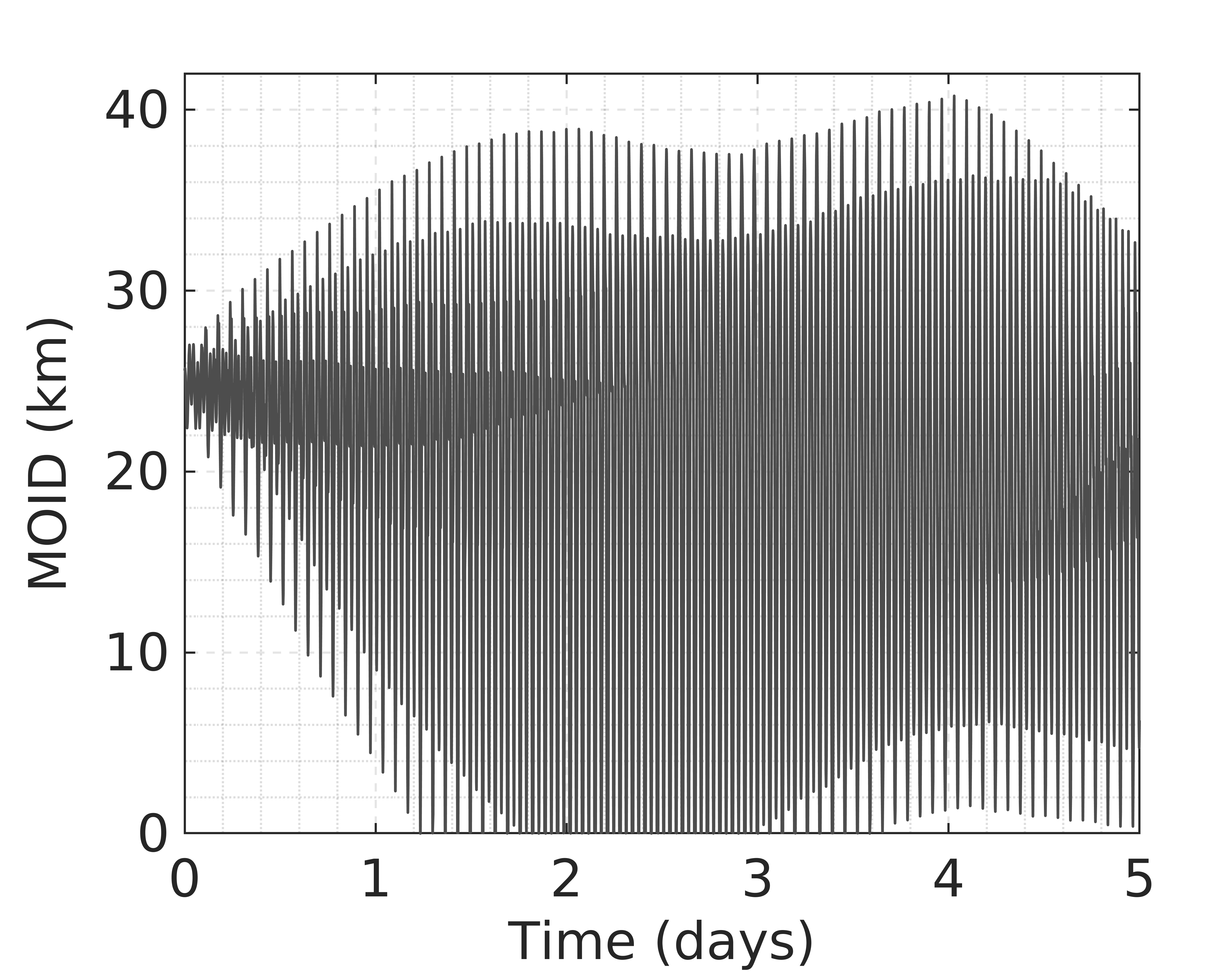}
  \includegraphics[width=0.49\textwidth]{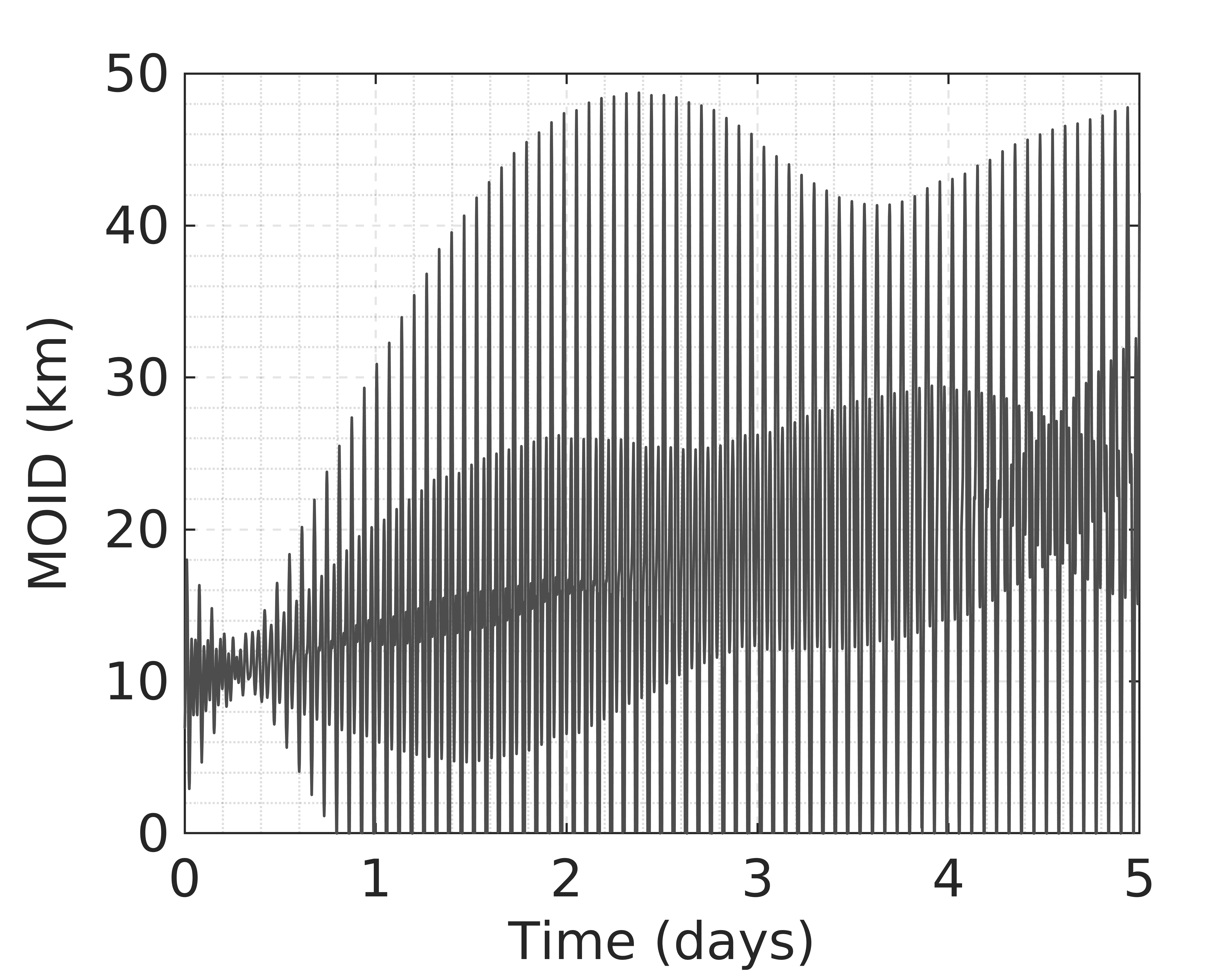}
  \caption{Keplerian osculating MOID for the two pairs of objects
    NORAD 41732, NORAD 42775 (left) and NORAD 41732, NORAD 41460
    (right).}
  \label{fig:MOID37} 
\end{figure}

Fig.~\ref{fig:MOID37} shows the evolution of the Keplerian osculating
MOID function computed at each integration time step for different
orbit pairs (the initial conditions are reported in
Table~\ref{tab:NORAD_data}) and using a high-fidelity dynamical
model. As it is evident, in both cases the osculating MOID function
goes to zero repeatedly. However, the numerically estimated true MOID
for the two cases yields values that are far from zero (22.97 km and
9.58 km, respectively). These examples suggest that Keplerian MOID
functions are unsuitable for the implementation of an efficient path
filter and that a non-Keplerian approach should be developed
instead. It is anticipated that the method developed in this work
estimates a minimum distance of 22.92 km and 8.57 km for the two
preceding test cases.

\section{Analytical expression of the orbit radius}
\label{sec:radius}

In a previous work \citep{SO_filter} related to the implementation of
a space occupancy filter (SO-filter), an analytical expression for the
evolution of the orbit radius was obtained as
\begin{equation}
  r\simeq\hat{a}\left(1-\hat{e}\cos(\hat{\theta}-\hat{\omega})\right)+
  \frac{J_{2}}{4\hat{a}}\left[\left(9+\cos 2\hat{\theta}\right)\sin^{2}\hat{i}-6\right],
  \label{eqn-hatr}
\end{equation}
where $J_{2}\approx1.082\times10^{-3}$ is the second Earth
gravitational harmonic, and $\hat{e}$, $\hat{\omega}$, $\hat{a}$,
$\hat{i}$, and $\hat{\theta}$ represent the mean values (averaged over
the mean anomaly) of the eccentricity, argument of perigee, semi-major
axis, inclination, and argument of latitude, respectively. The
analytical expressions of $\hat{e}$, $\hat{\omega}$ employed in
\cite{SO_filter} are based on the 1966 theory developed by Graham
E. Cook \citep{cook1966perturbations}, which accounts for a zonal
model with $J_{2}$ plus an arbitrary number of odd zonal terms while
$\hat{a}$ and $\hat{i}$ are kept constant.

Equation~\eqref{eqn-hatr} is simple enough to allow a fully analytical
implementation of an SO filter through the solution of a quartic
equation and provides sufficient accuracy to compute the critical
points of the orbit radius as required in this work (see
Section~\ref{sec:rbounds}).

However, a more refined expression for both the orbit radius and the
different mean orbital elements appearing in equation~\eqref{eqn-hatr}
is necessary for an accurate implementation of the orbit path filter
proposed here and will be dealt with in the following.

\subsection{Expression of the orbit radius including higher order eccentricity terms}

This section focuses on finding an approximation of the orbit radius
evolution with time that generalises equation~\eqref{eqn-hatr} by
including higher order effects in the mean eccentricity.  As done in
\citet{SO_filter}, the analytical expression for the radius is derived
by accounting for short-period effects depending linearly on the $J_2$
zonal harmonic term and neglecting the influence of higher order
harmonics. However, in order to make the formulation applicable up to
moderate eccentricities ($\hat{e}<0.1$), the next expansions are
considered up to order 3 in the eccentricity.

By using a classical result \citep[see][]{smart1953celestial}, the
orbit radius can be approximated as
\[
r\simeq a\left[1-e\cos M+\frac{1}{2}e^2(1-\cos 2M)+\frac{3}{8}e^3(\cos M-\cos 3M)\right].
\]
Each orbital element \textit{a}, \textit{e}, and \textit{M} is
replaced in the formula above with the sum of its mean ($\hat{a}$,
$\hat{e}$, $\hat{M}$) and short-periodic ($a_s$, $e_s$, $M_s$)
variation.  The expressions of $a_s$, $e_s$, $M_s$ derived by
\cite{Kozai} and \cite{Lyddane} can be written as
\[
a_s = \mathfrak{a}_s+J_2 O(\hat{e}),\qquad
e_s = \mathfrak{e}_s+J_2 O(\hat{e}),\qquad
\hat{e}M_s = \mathfrak{m}_s+J_2 O(\hat{e}),
\]
where
\begin{align}
  \mathfrak{a}_s &= \frac{3\kappa J_2}{2\hat{a}}\cos (2\hat{\nu}+2\hat{\omega}),\label{eq:asp}\\
  \mathfrak{e}_s &= \frac{J_2}{8\hat{a}^2}\left[7\kappa\cos(3\hat{\nu}+2\hat{\omega})+6(2-3\kappa)\cos\hat{\nu}
    +3\kappa\cos(\hat{\nu}+2\hat{\omega})\right],\label{eq:esp}\\
  \mathfrak{m}_s &= \frac{J_2}{8\hat{a}^2}\left[-7\kappa\sin(3\hat{\nu}+2\hat{\omega})-6(2-3\kappa)\sin\hat{\nu}
    +3\kappa\sin(\hat{\nu}+2\hat{\omega})\right],\label{eq:Msp}
\end{align}
with $\hat{\nu}$, $\hat{\omega}$ the mean values of the true anomaly
and argument of perigee, respectively, $\kappa=\sin^2\hat{i}$, and
$\hat{i}$ is the mean inclination. Then, by applying the following
approximation:
\[
\cos(jM) \simeq \cos(j\hat{M}) - jM_{s}\sin(j\hat{M}),\quad j=1,2,3,
\]
the orbit radius is given by
\[
\begin{aligned}
  r &\simeq (\hat{a}+a_s)\Bigl(1-(\hat{e}+e_s)(\cos\hat{M}-M_{s}\sin\hat{M})\\
  &\quad\, +\frac{1}{2}(\hat{e}+e_s)^2(1-\cos 2\hat{M}+2M_{s}\sin 2\hat{M})\\
  &\quad\, +\frac{3}{8}(\hat{e}+e_s)^3\left[\cos\hat{M}-M_{s}(\sin\hat{M}-3\sin3\hat{M})-\cos3\hat{M}\right]\Bigr).
\end{aligned}
\]
Neglecting the terms that contain $J_2\hat{e}^n$, with $n\ge 1$,
yields
\begin{equation}
  \begin{aligned}
    r &\simeq \hat{a}+\mathfrak{a}_s-\hat{a}\Bigl[(\hat{e}+\mathfrak{e}_s)\cos\hat{M}-\mathfrak{m}_s\sin\hat{M}
      -\frac{1}{2}\hat{e}^2(1-\cos 2\hat{M})\\
      &\quad\, -\frac{3}{8}\hat{e}^3(\cos\hat{M}-\cos3\hat{M})\Bigr].
  \end{aligned}
  \label{eq:r_Mm3}
\end{equation}
Finally, through the equation of the center
\citep[see][]{smart1953celestial}, one can write
\[
\hat{M}=\hat{\nu}-2\hat{e}\sin\hat{\nu}+\frac{3}{4}\hat{e}^2\sin 2\hat{\nu}
-\frac{1}{3}\hat{e}^3\sin 3\hat{\nu}+O(\hat{e}^4),
\]
so that
\begin{align}
  \cos\hat{M} &= \cos\hat{\nu}+2\hat{e}\sin^2\hat{\nu}-\frac{7}{2}\hat{e}^2\cos\hat{\nu}\sin^2\hat{\nu}
  +O(\hat{e}^3),\label{eq:cosMm}\\
  \cos 2\hat{M} &= \cos2\hat{\nu}+8\hat{e}\cos\hat{\nu}\sin^2\hat{\nu}+O(\hat{e}^2).\label{eq:cos2Mm}
\end{align}
Using~\eqref{eq:asp},~\eqref{eq:esp},~\eqref{eq:Msp},~\eqref{eq:cosMm},
and~\eqref{eq:cos2Mm} in equation~\eqref{eq:r_Mm3}, the final
expression of the orbit radius becomes
\begin{equation}
  \begin{aligned}
    r &\simeq \hat{a}\left(1-\hat{e}\cos(\hat{\theta}-\hat{\omega})\right)
    \left(1-\hat{e}^2\sin^2(\hat{\theta}-\hat{\omega})\right)\\
    &\quad\, +\frac{J_2}{4 \hat{a}}\left[\left(9+\cos 2\hat{\theta}\right)\sin^2\hat{i}-6\right],
  \end{aligned}
  \label{eq:radius}
\end{equation}
where the mean argument of latitude
$\hat{\theta}=\hat{\nu}+\hat{\omega}$ has been introduced.

In equation~\eqref{eq:radius} the quantities $\hat{a}$, $\hat{i}$ are
assumed to be constant, while $\hat{\theta}$, $\hat{e}$,
$\hat{\omega}$ vary with time. The evolutions of $\hat{e}$,
$\hat{\omega}$ are obtained from an extension of Graham E. Cook's
theory developed in the next section.

\subsection{Evolution of the eccentricity vector}

The theory formulated by Graham E. Cook
\citep[see][]{cook1966perturbations} analytically describes the motion
of the eccentricity vector under the effect of $J_2$ and an arbitrary
number of odd zonal harmonics on orbits of small eccentricity. Richard
A. Cook, in his 1992 paper \citep{cook1992}, expands upon this result
by including the impact of the remaining even harmonics by means of
the inclination and eccentricity functions introduced by \cite{Kaula}.
Although the latter theory represents an improvement of the former, it
still needs to be extended in order to be fully applicable to the
implementation of the proposed orbit path filter. The extension
involves the inclusion of higher order eccentricity terms.

In the following, the theory of Richard A. Cook is briefly summarized
for the reader's convenience before developing the required
eccentricity extension. For this purpose, Legendre polynomials are
employed as in \citet{cook1966perturbations}.  Let us adopt
dimensionless units of length and time, taking the Earth radius,
$R_{\oplus}$, as the reference length and $1/n_{\oplus}$ as the
reference time, with $n_{\oplus}$ indicating the mean motion of a
Keplerian circular orbit of radius $R_{\oplus}$.

The differential equations describing the evolution of the mean
eccentricity vector nodal components
\begin{equation}
  \xi = \hat{e}\cos\hat{\omega},\qquad 
  \eta = \hat{e}\sin\hat{\omega}
  \label{eq:xieta}
\end{equation}
are
\begin{equation}
  \begin{cases}
    \,\dot{\xi} = -k_{\xi}\eta+C,\\
    \,\dot{\eta} = k_{\eta}\xi,
  \end{cases}
  \label{eq:Cook}
\end{equation}
where $k_{\xi}$ and $k_{\eta}$ encompass the contribution due to the
even harmonics up to the first order in the eccentricity and are
defined by
\[
\begin{split}
  k_{\xi} &= -\sum_{n = 1}^{N}J_{2n}\hat{n}\left(\frac{R_{\oplus}}{\hat{a}}\right)^{2n}\left(\frac{2n(2n+1)}{2}P_{2n}(\cos\hat{i}\,)P_{2n}(0)\right.\\
  &\quad\, +\frac{(2n-1)(2n-2)(2n-2)!}{2(2n+2)!}P_{2n}^2(\cos\hat{i}\,)P_{2n}^2(0)\\
  &\quad\, -P_{2n}'(\cos\hat{i}\,)P_{2n}(0)\cot\hat{i}\Bigr),\\[0.5ex]
  k_{\eta} &= \sum_{n = 1}^{N}J_{2n}\hat{n}\left(\frac{R_{\oplus}}{\hat{a}}\right)^{2n}\left(-\frac{2n(2n+1)}{2}P_{2n}(\cos\hat{i}\,)P_{2n}(0)\right.\\
  &\quad\, +\frac{(2n-1)(2n-2)(2n-2)!}{2(2n+2)!} P_{2n}^2(\cos \hat{i}\,) P_{2n}^2(0)\\
  &\quad\, +P_{2n}'(\cos\hat{i}\,)P_{2n}(0)\cot\hat{i}\Bigr),
  \label{k_def}
\end{split}
\]
with $J_{2n}$ denoting the even zonal coefficient of degree $2n$.

\noindent
The contribution to $\dot{\xi}$, $\dot{\eta}$ due to the odd harmonics
up to the first order in the eccentricity appears in the constant term
$C$, which is given by
\[
C = \sum_{n = 1}^{N}J_{2n+1}\hat{n}\left(\frac{R_{\oplus}}{\hat{a}}\right)^{2n+1}\frac{n}{(2n+1)(n+1)}P_{2n+1}^1(\cos\hat{i}\,)P_{2n+1}^1(0),
\]
where $J_{2n+1}$ represents the odd zonal coefficient of degree
$2n+1$, $P_n$ is the Legendre polynomial of degree $n$, $P_n'$ its
derivative with respect to the inclination, and $P_n^s$ the associated
Legendre function of order $s$ and degree $n$. The solution of
system~\eqref{eq:Cook} is given by
\begin{equation}
  \begin{cases}
    \,\xi(\tau) = e_p \cos\left(k_{\xi}\sqrt{\frac{k_{\eta}}{k_{\xi}}}\tau + \alpha\right),\\[2ex]
    \,\eta(\tau) = \sqrt{\frac{k_{\eta}}{k_{\xi}}}e_p\sin\left(k_{\xi}\sqrt{\frac{k_{\eta}}{k_{\xi}}}\tau + \alpha\right) + e_f,
  \end{cases}
  \label{eq:sol_Cook}
\end{equation}
where $\tau$ denotes the non-dimensional time.  The additional term
$e_f$, known as \textit{frozen eccentricity}, is defined by
\[
e_f = \frac{C}{k_{\xi}}.
\]
Moreover, the amplitude $e_p$, which is known as \textit{proper
  eccentricity}, and the phase $\alpha$
depend on the initial conditions of the eccentricity and argument of
perigee:
\[
\begin{split}
  e_p &= \sqrt{\hat{e}_0^2\cos^2\hat{\omega}_0+\frac{k_{\xi}}{k_{\eta}}\left(\hat{e}_0\sin\hat{\omega}_0-e_f\right)^2},\\
  \cos \alpha &= \frac{\hat{e}_0\cos\hat{\omega}_0}{e_p},\quad\sin\alpha =
  \sqrt{\frac{k_{\xi}}{k_{\eta}}}\left(\frac{\hat{e}_0\sin\hat{\omega}_0-e_f}{e_p}\right).
\end{split}
\]
Solution~\eqref{eq:sol_Cook} corresponds to an ellipse of semiaxis
$e_p \sqrt{k_{\eta}/k_{\xi}}$ along the $\eta$ axis and semiaxis $e_p$
along the $\xi$ axis. Typically, the semi-major axis of this ellipse
aligns with the $\eta$-axis, except for orbits with inclinations very
close to the critical value.

\subsubsection{Including $J_2\hat{e}^2$ terms}
\label{sec:J2third}

\begin{figure}
  \centering
  \includegraphics[width=\textwidth]{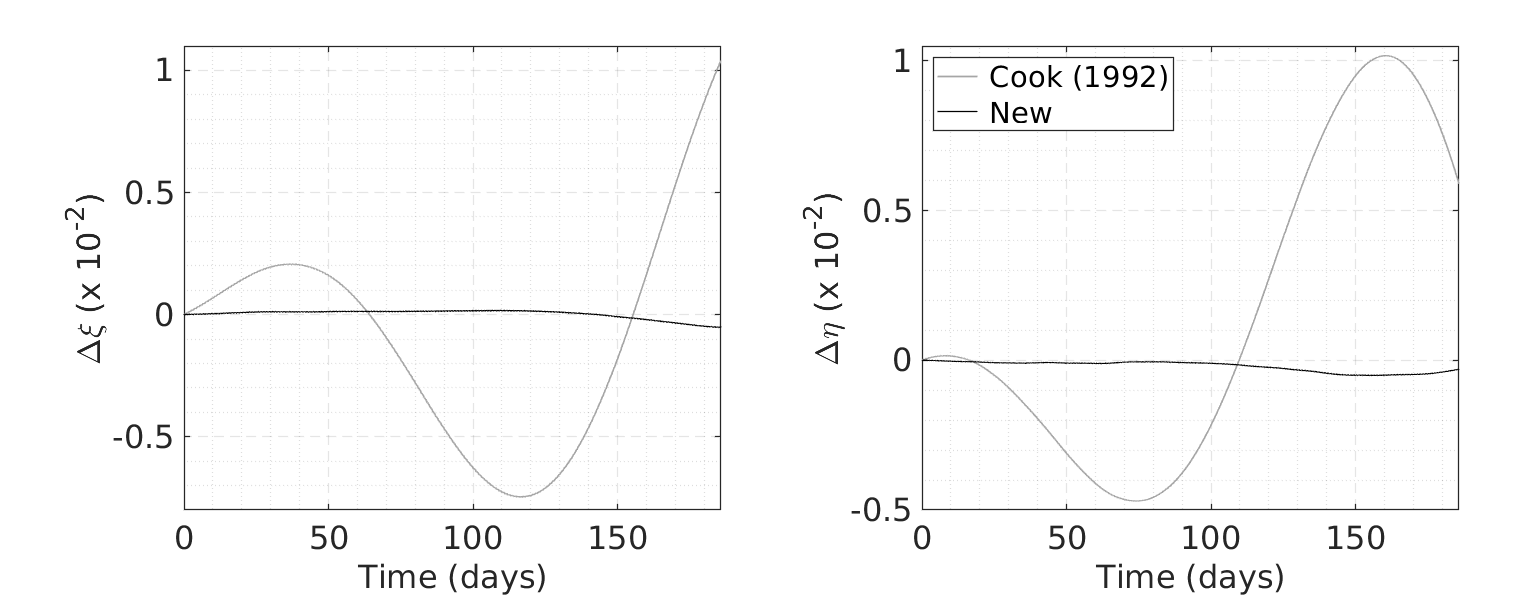}
  \caption{Evolution of the errors in the mean eccentricity vector
    nodal components computed from \citet{cook1992} and the improved
    theory leading to solution~\eqref{eq:sol_Cook_2}.}
  \label{fig:2nd_Cook}
\end{figure}

Both Graham and Richard Cook's analytical solutions neglect the effect
of nonlinear terms in the eccentricity. As it will be shown in the
following, the most important nonlinear terms are of the order of
$J_{2}\hat{e}^{2}$ and influence the phasing of the mean eccentricity
vector rotation in such a way that they cannot be neglected when
evaluating the distance between two orbits. A clarifying example of
this is summarized in Fig.~\ref{fig:2nd_Cook}, which illustrates the
evolution of the errors in the mean eccentricity vector nodal
components made by R. Cook's solution over one precession period of
the apsidal line for the catalogued object NORAD 29740. The initial
conditions, taken form the TLEs at epoch 2022~November~2, 16:22:53,
correspond to an initial eccentricity of $\approx\,$0.1. While the
amplitudes of oscillation of $\xi$ and $\eta$ are well-reproduced by
R. Cook's theory as compared to the numerical (averaged) solution, the
frequencies are smaller.

In order to improve the analytical solution, the contributions due to
the $J_2$ up to the third-order in the eccentricity are retained in
the expressions of $\dot{\xi}$, $\dot{\eta}$. After replacing $U'$ in
equations (3) and (4) of \citet{cook1966perturbations} with the
Earth's gravitational potential associated to the second harmonic
averaged over one orbit \citep[see][]{battin}
\[
U_2 = J_2(\hat{n}R_{\oplus})^2\left(\frac{1}{2}-\frac{3}{4}\sin^2\hat{i}\right)\bigl(1-\hat{e}^2\bigr)^{-3/2},
\]
where $\hat{n}=\sqrt{\mu/\hat{a}^3}$, and applying the aforementioned
approximation, one has
\[
\begin{cases}
  \,\dot{\xi} = -\left[k_{\xi}+2k\bigl(\xi^2+\eta^2\bigr)\right]\eta+C,\\
  \,\dot{\eta} = \left[k_{\eta}+2k\bigl(\xi^2+\eta^2\bigr)\right]\xi,
\end{cases}
\]
where
\[
k = 3J_2\hat{n}\left(\frac{R_{\oplus}}{\hat{a}}\right)^2\left(1-\frac{5}{4}\sin^2\hat{i}\right).
\]
These equations cannot be solved analytically. However, by assuming
that the squared eccentricity term $\xi^2+\eta^2$ is constant and
equal to its initial value $\hat{e}_0^2=\xi^2(0)+\eta^2(0)$, the
system becomes linear:
\begin{equation}
  \begin{cases}
    \,\dot{\xi} = -\kappa_{\xi}\eta+C,\\
    \,\dot{\eta} = \kappa_{\eta}\xi,
    \label{eq:J2_third_simpl}
  \end{cases}
\end{equation}
where 
\[
\begin{split}
  \kappa_{\xi} &= k_{\xi}+2k\hat{e}_0^2,\\
  \kappa_{\eta} &= k_{\eta}+2k\hat{e}_0^2.
\end{split}
\]
The proposed approximation is reasonable considering that the time
interval of interest in this work is 5 days, while the mean value of
the period of the line of apsides is typically around 150 days. The
solution of system~\eqref{eq:J2_third_simpl} can be expressed in the
same form as in~\eqref{eq:sol_Cook}, wherein $k_{\xi}$ and $k_{\eta}$
are replaced by $\kappa_{\xi}$ and $\kappa_{\eta}$, respectively:
\begin{equation}
  \begin{cases}
    \,\xi(\tau) = e_p\cos\left(\kappa_{\xi}\sqrt{\frac{\kappa_{\eta}}{\kappa_{\xi}}}\tau+\alpha\right),\\[1ex]
    \,\eta(\tau) = \sqrt{\frac{\kappa_{\eta}}{\kappa_{\xi}}}e_p\sin\left(\kappa_{\xi}\sqrt{\frac{\kappa_{\eta}}{\kappa_{\xi}}}\tau
    +\alpha\right)+e_f.
  \end{cases}
  \label{eq:sol_Cook_2}
\end{equation}

The evolution of the errors in the mean eccentricity vector nodal
components obtained using the third-order
correction~\eqref{eq:sol_Cook_2} is displayed in
Fig.~\ref{fig:2nd_Cook} with the grey dashed line. The new proposed
solution closely approximates the numerical one.

\subsection{Expression of the orbit radius including the evolution of the eccentricity vector}
Let us define the quantity
\begin{equation}
  \beta = \kappa_{\xi}\sqrt{\frac{\kappa_{\eta}}{\kappa_{\xi}}}\tau+\alpha,
  \label{eq:beta}
\end{equation}
which appears in solution~\eqref{eq:sol_Cook_2}. This angle shows a
\emph{slow} evolution as opposed to the \emph{fast} angle
$\hat{\theta}$ which was previously introduced in the
expression~\eqref{eq:radius} of the orbit radius.  Over a time
interval of 5 days, $\hat{\theta}$ varies cyclically in the range
$[0,2\pi)$ several times, while $\beta$ covers only a limited
arc. This significant difference in time scales allows us to assume
$\beta$ and $\hat{\theta}$ to be independent. Then, the orbital
distance can be considered as a function of $\beta$ and
$\hat{\theta}$. The expression of $r(\hat{\theta},\beta)$ is
obtained from~\eqref{eq:radius} by first using the definitions of
$\xi$ and $\eta$ given in~\eqref{eq:xieta} and then solution
\eqref{eq:sol_Cook_2}:
\begin{equation}
  \begin{aligned}
    r & \simeq \hat{a}\left(1-e_{p}\cos\beta\cos\hat{\theta}-\sqrt{\frac{\kappa_{\eta}}{\kappa_{\xi}}}e_{p}\sin\beta\sin\hat{\theta}
    -e_{f}\sin\hat{\theta}\right)\cdot\\
    & \quad\, \left[1-\left(e_{p}\cos\beta\sin\hat{\theta}-\sqrt{\frac{\kappa_{\eta}}{\kappa_{\xi}}}e_{p}\sin\beta\cos\hat{\theta}
      -e_{f}\cos\hat{\theta}\right)^{2}\right]\\
    & \quad\, +\frac{J_{2}}{4\hat{a}}\left[\left(9+\cos2\hat{\theta}\right)\sin^{2}\hat{i}-6\right].
    \label{eq:radius2}
  \end{aligned}
\end{equation}

\section{The orbit path filter}
\label{sec:SOP_filter}

This section describes the algorithm developed for the orbit path
filter, which is formulated as an optimization problem.  For each
object of a given pair, the proposed procedure estimates the minimum
and maximum values of the orbit radius among those attained in a given
time interval $[t_0,t_{\rm{f}}]$ when the object is located at the
mutual nodes. From the two minima and two maxima, relative to the two
mutual nodes, a filtering criterion is formulated.

The absolute minimum and maximum of the orbit radius
$r(\hat{\theta},\beta)$ are searched in two compact domains
\begin{equation}
  \mathcal{D} = \mathcal{T}\times\mathcal{B},\qquad
  \mathcal{D}^* = \mathcal{T}^*\times\mathcal{B}
  \label{eq:D}  
\end{equation}
as described in Section~\ref{sec:rbounds}. The interval for the
variable $\beta$ is
\[
\mathcal{B} = [\beta_{\rm min},\beta_{\rm max}],
\]
where the extrema $\beta_{\rm min}$, $\beta_{\rm max}$ are given by
setting $\tau=t_0, t_{\rm f}$, respectively, in~\eqref{eq:beta}. The
computation of the domains $\mathcal{T}$, $\mathcal{T}^*$ for the
variable $\hat{\theta}$ is described in Section~\ref{sec:opdom}. It is
based on the assumption that if the mutual inclination between the
orbital planes of the two objects is sufficiently large, the MOID is
achieved by two points that are close to one pair of mutual nodes
\citep{Gronchi02}.

In Section~\ref{sec:filtering} it is explained how the filter employs
the minimum and maximum values of the orbit radius in each of the two
domains $\mathcal{D}$, $\mathcal{D}^*$ and for each of the two objects
under consideration to evaluate whether a possible conjunction event
could occur between them in $[t_0,t_{\rm f}]$.

\subsection{Computation of the intervals $\mathcal{T}$, $\mathcal{T}^*$}
\label{sec:opdom}

The location of the mutual nodes is predicted by taking into account
the influence of the geopotential on the longitudes of the nodes. In
particular, the secular effect due to the zonal harmonics $J_2$,
$J_3$, $J_4$ derived in \cite{Kozai62} has been implemented for the
time evolution of $\hat{\Omega}$.

\begin{figure}
  \centering
  \includegraphics[width=0.5\textwidth]{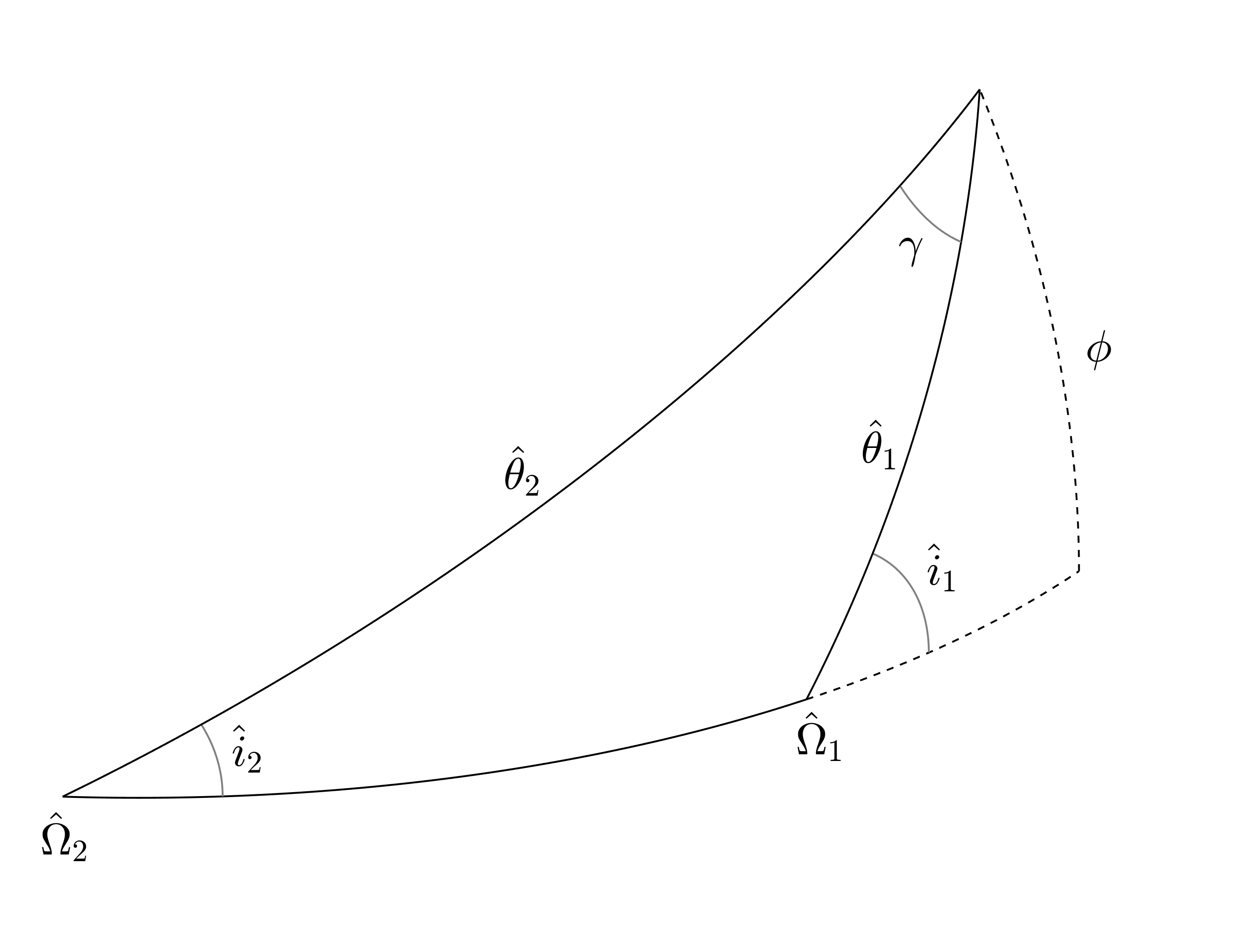}
  \caption{Spherical triangles generated from the ascending nodes and
    one pair of mutual nodes of two trajectories. The mutual
    inclination $\gamma$ is also indicated.}
  \label{fig:lat_tri} 
\end{figure}

For convenience, all orbits are considered direct (i.e.,
$\hat{i}\in[0,\pi/2])$. If $\hat{i}>\pi/2$, then $\hat{i}$ is replaced
by $\pi-\hat{i}$ and $\hat{\Omega}$ by $\pi+\hat\Omega$
($\bmod{\,2\pi}$). Moreover, it is not restrictive to assume that
$\hat{i}_1>\hat{i}_2$.  The latitude of one of the two pairs of mutual
nodes is given by\footnote{The latitude of the other pair is equal to
  $-\phi$.} (see Fig.~\ref{fig:lat_tri})
\[
\phi = \arcsin\left(\frac{\sin\hat{i}_1\sin\hat{i}_2\sin\Delta\hat{\Omega}}{\sin\gamma}\right),
\]
where $\hat{i}_1$, $\hat{i}_2$ are the mean inclinations of the two
orbital planes, $\Delta\hat{\Omega}=\hat{\Omega}_2-\hat{\Omega}_1$,
and the mutual inclination $\gamma$, which is defined as the angle
between the directions orthogonal to the orbital planes, is found from
relation
\[
\gamma = \arccos\Bigl(\cos\hat{i}_1\cos\hat{i}_2+\sin\hat{i}_1\sin\hat{i}_2\cos\Delta\hat{\Omega}\Bigr).
\]

\begin{figure}[h]
  \centering
  \includegraphics[width=0.9\textwidth]{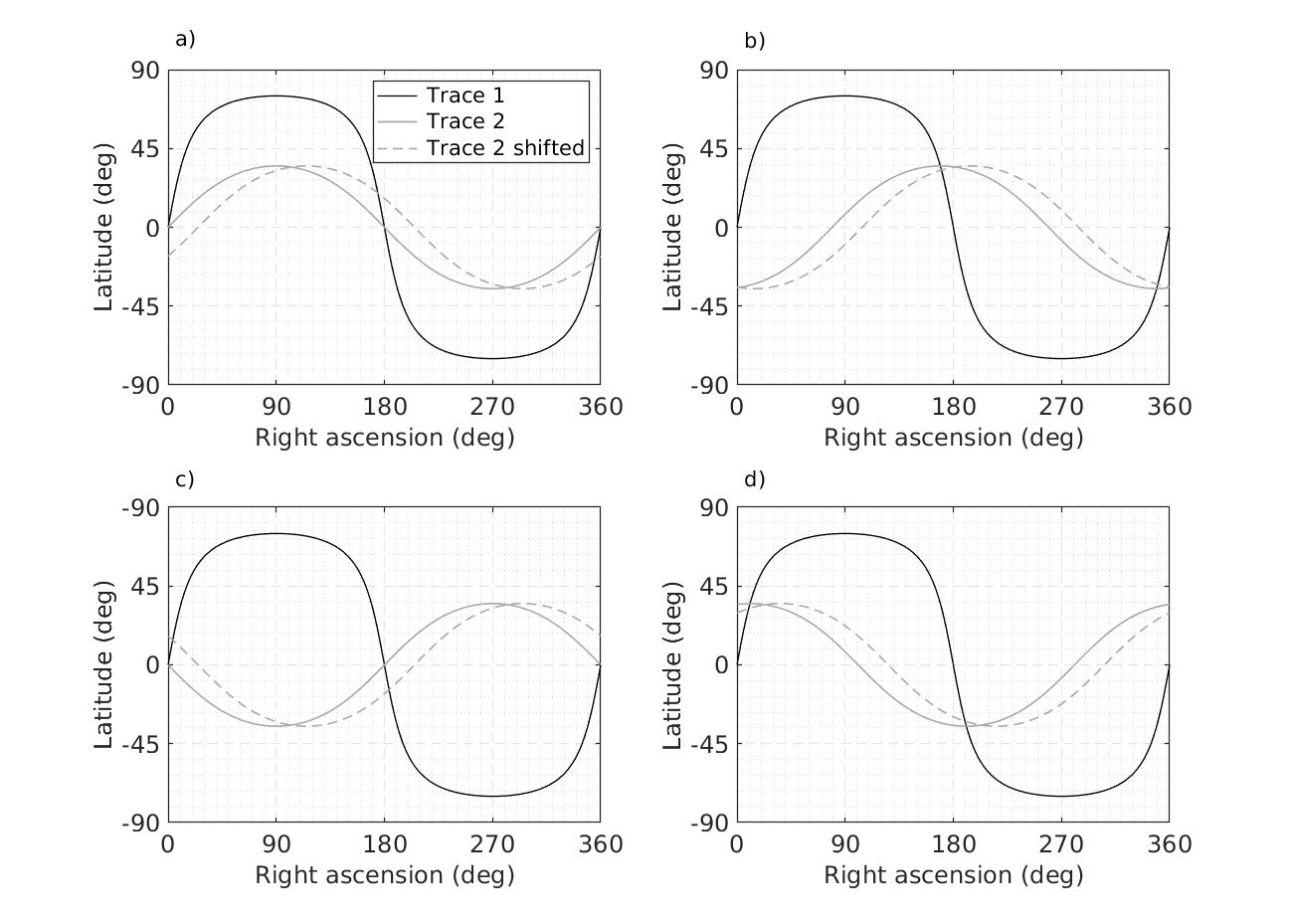}
  \caption{Orbital traces (``Trace 1'' and ``Trace 2'') of two
    ellipses (their orbital elements are specified in the text) in
    four different configurations: a) $\Delta\hat{\Omega}=0$, b)
    $\Delta\hat{\Omega}=\alpha$, c) $\Delta\hat{\Omega}=\pi$, d)
    $\Delta\hat{\Omega}=2\pi-\alpha$, where
    $\Delta\hat{\Omega}=\hat{\Omega}_2-\hat{\Omega}_1$ and $\alpha$ is
    defined in equation~\eqref{eq:alpha}. ``Trace 2 shifted'' is
    obtained by adding $5\pi/36$ to $\hat{\Omega}_2$.}
  \label{fig:trace} 
\end{figure}

\noindent
If $\hat i_1\hat i_2 \ne0$, the arguments of latitude of the mutual
nodes are computed from
\begin{equation}
  \sin\hat{\theta}_k = \frac{\sin\phi}{\sin\hat i_k},\qquad k=1,2,
  \label{eq:theta}
\end{equation}
which admits two solutions in $[0,2\pi]$. The ambiguity can be solved
by looking at the projection of the orbit over a non-rotating Earth in
the inertial reference frame, here named \emph{trace}. The relative
displacement of the traces of two (non-coplanar) ellipses as
$\Delta\hat{\Omega}$ is varied in the range $[0,2\pi]$ is shown in
Fig.~\ref{fig:trace}. The motion along the traces occurs from left to
right. The curve labelled ``Trace 1'' refers to the same trajectory in
the four scenarios, which is given by $\hat{a}_1=7,300$ km,
$\hat{e}_1=0.002$, $\hat{i}_1=75$ deg, $\hat{\omega}_1=90$ deg,
$\hat{\Omega}_1=0$ deg. The curve ``Trace 2'' refers to the
trajectory defined by $\hat{a}_2=7,800$ km, $\hat{e}_2=0.008$,
$\hat{i}_2=35$ deg, $\hat{\omega}_2=0$ deg, while $\hat{\Omega}_2$
takes four different values. The third trace (``Trace 2 shifted'')
displayed in Fig.~\ref{fig:trace} is generated from the same
trajectory relative to Trace 2 by changing $\hat{\Omega}_2$ of a small
amount. Note that the value $\hat{\Omega}_1=0$ is chosen only for
convenience and the following analysis is valid for any value of
$\hat{\Omega}_1$.

\begin{figure}[h]
  \centering
  \includegraphics[width=0.5\textwidth]{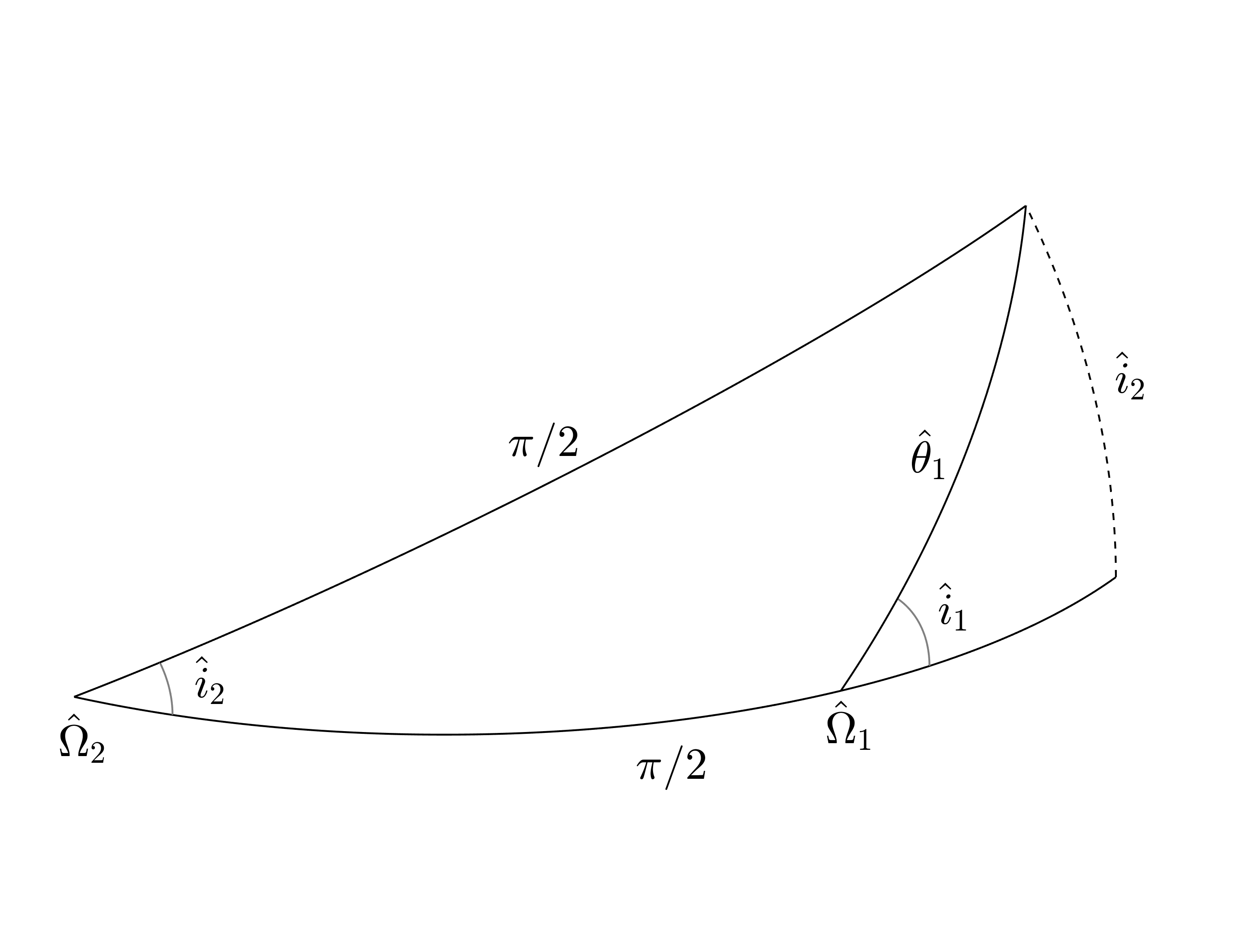}
  \caption{Spherical triangles generated from the ascending nodes and
    one pair of mutual nodes of two trajectories in the particular
    case $\phi=\hat{i}_2$.}
  \label{fig:tri} 
\end{figure}

Four particular configurations can be identified: the mutual nodes lie
on the equatorial plane ($\phi=0$), so that either
$\Delta{\hat{\Omega}}=0$ (Fig.~\ref{fig:trace}a) or
$\Delta{\hat{\Omega}}=\pi$ (Fig.~\ref{fig:trace}c); the latitudes of
the two pairs of mutual nodes are equal to the maximum and minimum
latitudes reached along ``Trace 2'' ($\phi=\pm\hat{i}_2$, see
Figs.~\ref{fig:trace}b and~\ref{fig:trace}d). The latter condition is
realized by two values of $\Delta{\hat{\Omega}}$ which are computed as
follows. Consider the configuration shown in Fig.~\ref{fig:tri}, which
refers to Fig.~\ref{fig:trace}d, and in particular to the intersection
point between ``Trace 1'' and ``Trace 2'' with the smaller right
ascension. Noting that $\phi=\hat{i}_2$ and $\hat{\theta}_2=\pi/2$,
the argument of latitude of the mutual node along the trajectory of
Trace 1 is given by
\begin{equation}
  \hat{\theta}_1^* = \arcsin\left(\frac{\sin\hat{i}_2}{\sin\hat{i}_1}\right),
  \label{eq:theta_ref}
\end{equation}
and the value taken by $\hat{\Omega}_1-\hat{\Omega}_2\pmod{2\pi}$,
here denoted by $\alpha$, is computed from the formula
\begin{equation}
  \alpha=\frac{\pi}{2}-\arctan\left(\tan\hat{\theta}_1^* \cos\hat{i}_1\right).
  \label{eq:alpha}
\end{equation}
Then, it results $\Delta\hat{\Omega}=2\pi-\alpha$. In a similar way,
it is possible to see that the configuration of traces 1 and 2 in
Fig.~\ref{fig:trace}b is realized by $\Delta\hat{\Omega}=\alpha$, with
$\alpha$ and $\hat{\theta}_1^*$ still defined as in~\eqref{eq:alpha}
and~\eqref{eq:theta_ref}.

Therefore, as $\Delta\hat{\Omega}$ is increased from $0$ to $2\pi$,
one realizes through Fig.~\ref{fig:trace} that the arguments of
latitude of the pair of mutual nodes in the Northern hemisphere (i.e.,
$\phi\in[0,\pi/2]$) belong to the intervals reported in
Table~\ref{tab:intervals}.

\renewcommand{\arraystretch}{1.3}
\begin{table}
  \centering 
  \begin{tabular}{lcc}
    \hline 
    $\Delta\hat{\Omega}$ & $\hat\theta_1$ & $\hat\theta_2$ \\
    \hline 
    $I_a=[0,\alpha]$ & $[\pi/2,\pi]$ & $[\pi/2,\pi]$ \\
    $I_b=[\alpha,\pi]$ & $[\pi/2,\pi]$ & $[0,\pi/2]$ \\
    $I_c=[\pi,2\pi-\alpha]$ & $[0,\pi/2]$ & $[\pi/2,\pi]$ \\
    $I_d=[2\pi-\alpha,2\pi]$ & $[0,\pi/2]$ & $[0,\pi/2]$ \\
    \hline
  \end{tabular}
  \caption{Ranges of the arguments of latitude of the pair of mutual
    nodes in the Northern hemisphere as $\Delta\hat{\Omega}$ varies in
    $[0,2\pi]$.}
  \label{tab:intervals}
\end{table}

In the time span $[t_0,t_{\rm{f}}]$ the quantity $\Delta\hat{\Omega}$
varies monotonically from
\[
\Delta\hat{\Omega}_0 = \hat{\Omega}_2(t_0)-\hat{\Omega}_1(t_0)
\]
to
\[
\Delta\hat{\Omega}_{\rm{f}} = \hat{\Omega}_2(t_{\rm{f}})-\hat{\Omega}_1(t_{\rm{f}}),
\]
where, the final values
$\hat{\Omega}_1(t_{\rm{f}}),\hat{\Omega}_2(t_{\rm{f}})$ will differ by
a few degrees from the initial ones, since $t_{\rm{f}}=5$~days. Let
us restrict $\Delta\hat{\Omega}_0$, $\Delta\hat{\Omega}_{\rm{f}}$ to
$[0,2\pi]$ and introduce
\[
I_\Omega=
\begin{cases}
  [\Delta\hat{\Omega}_0,\Delta\hat{\Omega}_{\rm{f}}] & {\rm if}\,\,\Delta\hat{\Omega}_{\rm{f}}>\Delta\hat{\Omega}_0,\\
  [\Delta\hat{\Omega}_0,2\pi)\cup[0,\Delta\hat{\Omega}_{\rm{f}}] & {\rm if}\,\,\Delta\hat{\Omega}_0>\Delta\hat{\Omega}_{\rm{f}}.
\end{cases}
\]
There are three different cases to discuss.

If $I_\Omega\subset I_k$, $k\in\{a,b,c,d\}$, where the intervals $I_k$
are defined in Table~\ref{tab:intervals}, the arguments of latitude of
the pair of mutual nodes in the Northern hemisphere vary monotonically
for $\tau\in[t_0,t_{\rm{f}}]$. Then, one can set
\begin{equation}
  \begin{aligned}
    \hat{\theta}_{{\rm min},k} &= \min\left(\hat{\theta}_k(t_0),\hat{\theta}_k(t_{\rm{f}})\right),\\
    \hat{\theta}_{{\rm max},k} &= \max\left(\hat{\theta}_k(t_0),\hat{\theta}_k(t_{\rm{f}})\right),
  \end{aligned}
  \label{eq:theta_int}
\end{equation}
where $\hat{\theta}_k(t_0)$, $\hat{\theta}_k(t_{\rm{f}})$ are computed
through relation~\eqref{eq:theta} with $\phi\in(0,\pi/2]$ and
  Table~\ref{tab:intervals}. The intervals $\mathcal{T}_k$,
  $\mathcal{T}^*_k$ are given by
  \begin{equation}
    \begin{aligned}
      \mathcal{T}_k &=
      [\hat{\theta}_{{\rm min},k},\hat{\theta}_{{\rm max},k}],\\
      \mathcal{T}^*_k &= [\hat{\theta}_{{\rm min},k}+\pi,\hat{\theta}_{{\rm max},k}+\pi].
    \end{aligned}
    \label{eq:Ith}
  \end{equation}

If instead at some time $\bar{t}$ the latitudes of the mutual nodes
reach the maximum ($\hat{i}_2$) and minimum ($-\hat{i}_2$) values,
that is, $\alpha\in I_\Omega$ or $2\pi-\alpha\in I_\Omega$, then the
time derivative of $\hat{\theta}_1$ will vanish at $\bar{t}$. Thus,
$\hat{\theta}_1$ will increase in $[t_0,\bar{t})$ and decrease in
  $(\bar{t},t_{\rm{f}}]$, while $\hat{\theta}_2$ will vary
monotonically. As a consequence, $\hat{\theta}_{{\rm min},2}$,
$\hat{\theta}_{{\rm max},2}$ can be computed
from~\eqref{eq:theta_int}, and
\begin{equation}
  \begin{cases}
    \hat{\theta}_{{\rm min},1}=\pi-\hat{\theta}_1^* & {\rm if}\,\,\alpha\in I_\Omega,\\
    \hat{\theta}_{{\rm max},1}=\hat{\theta}_1^* & {\rm if}\,\,2\pi-\alpha\in I_\Omega,
  \end{cases}
  \label{eq:thminmax1}
\end{equation}
where $\hat{\theta}_1^*$ is defined in~\eqref{eq:theta_ref}. Moreover,
$\hat{\theta}_{{\rm max},1}$ (if $\alpha\in I_\Omega$),
$\hat{\theta}_{{\rm min},1}$ (if $2\pi-\alpha\in I_\Omega$) are given
by~\eqref{eq:theta_int}. Also in this case $\mathcal{T}_k$,
$\mathcal{T}^*_k$ are defined as shown in~\eqref{eq:Ith}.
 
Finally, if the two pairs of mutual nodes cross the equatorial plane,
that is, if $0\in I_\Omega$ or $\pi\in I_\Omega$, the arguments of
latitude of the mutual nodes vary monotonically for
$\tau\in[t_0,t_{\rm{f}}]$. In this case, the intervals for
$\hat{\theta}$ result
\begin{equation}
  \begin{aligned}
    \mathcal{T}_k &= [\hat{\theta}_{{\rm max},k}+\pi,2\pi)\cup[0,\hat{\theta}_{{\rm min},k}],\\
      \mathcal{T}^*_k &= [\hat{\theta}_{{\rm max},k},\hat{\theta}_{{\rm min},k}+\pi],
  \end{aligned}
  \label{eq:Ith2}
\end{equation}
where, for each object, $\hat{\theta}_{{\rm min},k}$,
$\hat{\theta}_{{\rm max},k}$ are computed from~\eqref{eq:theta_int}.

The intervals $\mathcal{T}_k$, $\mathcal{T}^*_k$ can be inflated
through the application of an \emph{angular buffer} (see
Section~\ref{sec:Int_buf}).

\subsection{Orbit radius bounds}
\label{sec:rbounds}

This section presents an efficient method for determining the minimum
and maximum values taken by the orbit radius $r(\hat{\theta},\beta)$
(see~\ref{eq:radius2}) at the mutual nodes in each domain
$\mathcal{D}$, $\mathcal{D}^*$ (see~\ref{eq:D}).\footnote{In this
  section, the subscript $k=1,2$ which refers to the two objects of
  the considered pair is dropped to simplify the exposition.}

First, the critical points of the function $r(\hat{\theta},\beta)$ are
computed on its domain of definition $[0,2\pi]\times\mathbb{R}$. Since
their computation would considerably slow down the SOP-filter, a
simpler approach is adopted\footnote{The critical points of
  $r(\hat{\theta},\beta)$ are solutions of a system of two equations,
  each involving a bivariate trigonometric polynomial, which can be
  transformed to an ordinary polynomial by a change of
  variables. Then, by resultant theory \citep{Cox} a univariate
  polynomial can be obtained such that all its roots represent one
  component of all the solutions of the system made by the two
  ordinary polynomials. In our case, the degree of such polynomial is
  72 and the computation of its coefficients and its roots requires a
  complex procedure.}. Four critical points can be easily found:
$(\pi/2,\pi/2)$, $(\pi/2,3\pi/2)$, $(3\pi/2,\pi/2)$,
$(3\pi/2,3\pi/2)$. Additional critical points to these four can appear
only if the frozen and proper eccentricities, and accordingly
$\hat{e}$, are $\ll 1$. By neglecting the nonlinear terms in the
eccentricity in the expression of $r$ given in~\eqref{eq:radius2}, the
resulting function still admits the four critical points reported
above and allows for an explicit computation of the additional
critical points when they exist as described in \cite{SO_filter},
where Graham Cook's theory is employed (i.e.,
$\kappa_\xi=\kappa_\eta$). The corresponding critical values of $r$
are always calculated using equation~\eqref{eq:radius2}.

If an absolute minimum/maximum point belongs to $\mathcal{D}$
($\mathcal{D}^*$), then, the corresponding value of $r$ is computed
through equation~\eqref{eq:radius2}. Otherwise, one has to search for
the minimum/maximum of $r$ on the border of $\mathcal{D}$
($\mathcal{D}^*$) through the procedure described in
Section~\ref{sec:borderD}, check whether there are local
minimum/maximum points of $r$ that belong to $\mathcal{D}$
($\mathcal{D}^*$), and in this case, select the smallest/largest value
of $r$.

Let $r_{{\rm min},k}$, $r_{{\rm max},k}$ and $r^*_{{\rm min},k}$,
$r^*_{{\rm max},k}$ be the minimum and maximum orbit radii in the
domains $\mathcal{D}_k$ and $\mathcal{D}^*_k$, respectively, where
$k=1,2$ refers to an object of the pair.

\subsubsection{Bounds of the orbit radius on the border of $\mathcal{D}$ and $\mathcal{D}^*$}
\label{sec:borderD}

Let us denote by $\partial\mathcal{D}$, $\partial\mathcal{D}^*$ the
borders of $\mathcal{D}$, $\mathcal{D}^*$, respectively, and introduce
the functions
\begin{align*}
r_{\beta_*}(\hat{\theta}) &= r(\hat{\theta},\beta_*),\\
r_{\hat{\theta}_*}(\beta) &= r(\hat{\theta}_*,\beta),
\end{align*}
where $r(\hat{\theta},\beta_*)$, $r(\hat{\theta}_*,\beta)$ are two
trigonometric polynomials obtained from $r(\theta,\beta)$
in~\eqref{eq:radius2} by replacing $\beta$, $\hat{\theta}$ with some
constants $\beta_*$, $\hat{\theta}_*$, respectively. For both
$\partial\mathcal{D}$ and $\partial\mathcal{D}^*$ one has to take
$\beta_*=\beta_{\rm min},\beta_{\rm max}$. Moreover, for
$\partial\mathcal{D}$, either
$\hat{\theta}_*=\hat{\theta}_{\rm{min}},\hat{\theta}_{\rm{max}}$
(see~\ref{eq:Ith}) or
$\hat{\theta}_*=\hat{\theta}_{\rm{min}}+\pi,\hat{\theta}_{\rm{max}}+\pi$
(see~\ref{eq:Ith2}) and for $\partial\mathcal{D}^*$, either
$\hat{\theta}_*=\hat{\theta}_{\rm{min}}+\pi,\hat{\theta}_{\rm{max}}+\pi$
(see~\ref{eq:Ith}) or
$\hat{\theta}_*=\hat{\theta}_{\rm{max}},\hat{\theta}_{\rm{min}}+\pi$
(see~\ref{eq:Ith2}). A point $(\hat{\theta}_*,\beta_*)$ defines a
\emph{vertex} of the domain. 

The critical points of $r_{\beta_*}(\hat{\theta})$,
$r_{\hat{\theta}_*}(\beta)$ are solutions of the equations
\begin{equation}
  \frac{d r_{\beta_*}}{d\hat{\theta}}=0,\qquad
  \frac{d r_{\hat{\theta}_*}}{d\beta}=0,
  \label{eq:dr_dth_db}
\end{equation}
where
\begin{align*}
  \frac{d r_{{\beta}_*}}{d\hat{\theta}} &= B_0+\sin\hat{\theta}\sum_{m=0}^{2}S_m\sin^m\hat{\theta}
  +\cos\hat{\theta}\sum_{m=0}^{2}C_m\sin^m\hat{\theta},
\end{align*}
for some coefficients $B_0(\beta_*)$, $S_m(\beta_*)$, $C_m(\beta_*)$,
and
\begin{align*}
  \frac{d r_{\hat{\theta}_*}}{d\beta} &= b_0+\sin\beta\sum_{m=0}^{2}s_m\sin^m\beta
  +\cos\beta\sum_{m=0}^{2}c_m\sin^m\beta,
\end{align*}
for some coefficients $b_0(\hat{\theta}_*)$, $s_m(\hat{\theta}_*)$,
$c_m(\hat{\theta}_*)$. Using relations
\begin{align*}
  \sin\hat{\theta} &= \frac{2x}{1+x^2},\quad
  \cos\hat{\theta}=\frac{1-x^2}{1+x^2},\\
  \sin\beta &= \frac{2y}{1+y^2},\quad
  \cos\beta = \frac{1-y^2}{1+y^2},
\end{align*}
equations~\eqref{eq:dr_dth_db} can be transfomed in two equations
involving the ordinary polynomials
\begin{equation}
q(x) = \sum_{n=0}^{6}q_n x^n = 0,\qquad
p(y) = \sum_{n=0}^{6}p_n y^n = 0,
\label{eq:pq}
\end{equation}
where the coefficients $q_n(\beta_*)$, $p_n(\hat{\theta}_*)$ are
provided in Appendix~\ref{ap:coef}.  For each real root $\bar{x}$ of
$q(x)$ there exists a unique critical point of $r_{\beta_*}$, given by
\[
\hat{\theta}^c = 2\arctan\bar{x}\pmod{2\pi}.
\]
Instead, for each real root $\bar{y}$ of $p(y)$ there exist infinite
critical points of $r_{\hat{\theta}_*}$, given by
\[
\beta^c_h = 2\arctan\bar{y}+2h\pi,\qquad h\in\mathbb{Z}.
\]
Consider $\partial\mathcal{D}$, the following discussion being the
same for $\partial\mathcal{D}^*$.  If for the values of $\beta_*$ and
$\hat{\theta}_*$ that define $\partial\mathcal{D}$, none of the points
$(\hat{\theta}^c,\beta_*)$, $(\hat{\theta}_*,\beta^c_h)$ lie on
$\partial\mathcal{D}$, then necessarily the minimum and maximum of $r$
on $\partial\mathcal{D}$ are reached at two vertices of $\mathcal{D}$.
Otherwise, one has to find the smallest and largest values of $r$
among those (if any) computed at the vertices of $\mathcal{D}$ and at
the points $(\hat{\theta}^c_h,\beta_*)$, $(\hat{\theta}^*,\beta^c)$
that lie on $\partial\mathcal{D}$.

Finally, if either $\pi\in \mathcal{T}$ or $\pi\in \mathcal{B}$, the
values of the orbit radius at the points on $\partial\mathcal{D}$ with
$\hat{\theta}=\pi$ or $\beta=\pi$ are compared with the minimum and
maximum orbit radii found by the procedure described above.

\subsection{Filter output}
\label{sec:filtering}

Consider a pair of space objects and introduce the following intervals
($k=1,2$):
\[
\mathcal{R}_k = \bigl[r_{{\rm min},k},r_{{\rm max},k}\bigr],\qquad
\mathcal{R}^*_k = \bigl[r^*_{{\rm min},k},r^*_{{\rm max},k}\bigr].
\]
If 
\begin{equation}
  \mathcal{R}_1\cap\mathcal{R}_2 = \emptyset
  \quad\land\quad
  \mathcal{R}^*_1\cap\mathcal{R}^*_2 = \emptyset,
  \label{eq:neg}
\end{equation}
the pair is flagged as \emph{negative} and is not considered for
further investigation, as a conjunction between the two objects is
deemed impossible. Otherwise, if
\begin{equation}
  \mathcal{R}_1\cap\mathcal{R}_2\ne\emptyset
  \quad\lor\quad
  \mathcal{R}^*_1\cap\mathcal{R}^*_2\ne\emptyset,
  \label{eq:pos}
\end{equation}
it is flagged as \emph{positive}, thus requiring more in-depth
analysis.

The intervals $\mathcal{R}_k$, $\mathcal{R}^*_k$ can be enlarged
through the application of a \emph{radial buffer} (see
Section~\ref{sec:R_buf}).

\section{Results}
\label{sec:Results}

The SOP-filter algorithm, developed in Section~\ref{sec:SOP_filter},
has been tested and validated using the same dataset as well as the
same high-fidelity dynamical model employed in \cite{SO_filter}.  From
the publicly available Two-Line Element sets (TLEs), 16,951 orbits
were extracted and then processed using the SGP4 theory
\citep{vallado2006revisiting}. The dataset is obtained by removing
orbits with an eccentricity higher than 0.1 and an apogee radius
exceeding 40,000 km. Moreover, orbits with inclinations smaller than
0.06 deg are also excluded (21 orbits from the sample).

The dynamical model includes a $23\times 23$ geopotential along with
luni-solar third-body perturbations, and it accounts for Earth's geoid
precession, nutation, and polar motion effects. Earth orientation, the
values of the harmonic coefficients, and the positions of the Sun and
Moon are obtained from the corresponding SPICE kernels. Note that the
SOP-filter is based on a model where only the zonal harmonics are
taken into account and Earth's polar axis is aligned with the $z$-axis
of the J2000 inertial frame.

The orbits of the dataset were propagated to a common epoch $t_0$
(11/02/2022, 09:18:20) and then processed by these two filters in
sequence: the SO-filter developed in \cite{SO_filter} and the
SOP-filter proposed in this work. The time window $t_{\rm f}-t_{\rm
  0}$ was set equal to 5 days. The SO-filter passes 32,474,006 out of
the 143,659,725 input pairs.  The SOP-filter is applied only to the
pairs with a mutual inclination in the range $[10,170]$~deg. Pairs not
satisfying this condition amount to 1,637,190 (about 5\%) and are
flagged as positive, thus requiring further analysis. The remaining
30,836,816 pairs are processed by the
SOP-filter. Table~\ref{tab:real_res} summarises some information about
the numerical test. In particular, it reports the \emph{real} number
of pairs that could potentially generate a collision according to
condition~\eqref{eq:pos}, as computed from numerical propagations.

\begin{figure}
  \centering
  \includegraphics[width=0.49\textwidth]{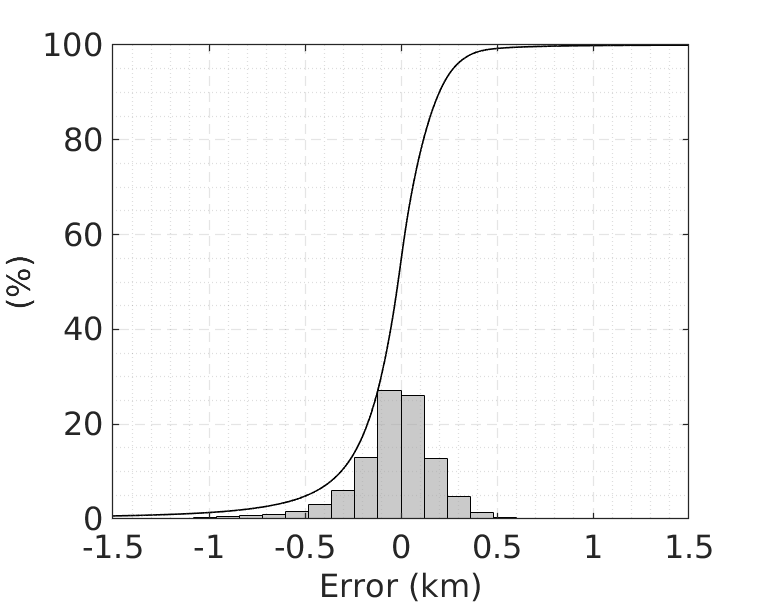}
  \caption{Histogram and cumulative density function of the errors in
    the minimum and maximum values of orbit radius inside the domains
    $\mathcal{D}$, $\mathcal{D}^*$ (see Section~\ref{sec:rbounds}) for
    the input pairs of the SOP-filter. A negligible percentage of
    errors is outside the interval $[-1.5,1.5]$~km.}
  \label{fig:GE_hist} 
\end{figure}

\begin{figure}
  \centering
  \includegraphics[width=0.49\textwidth]{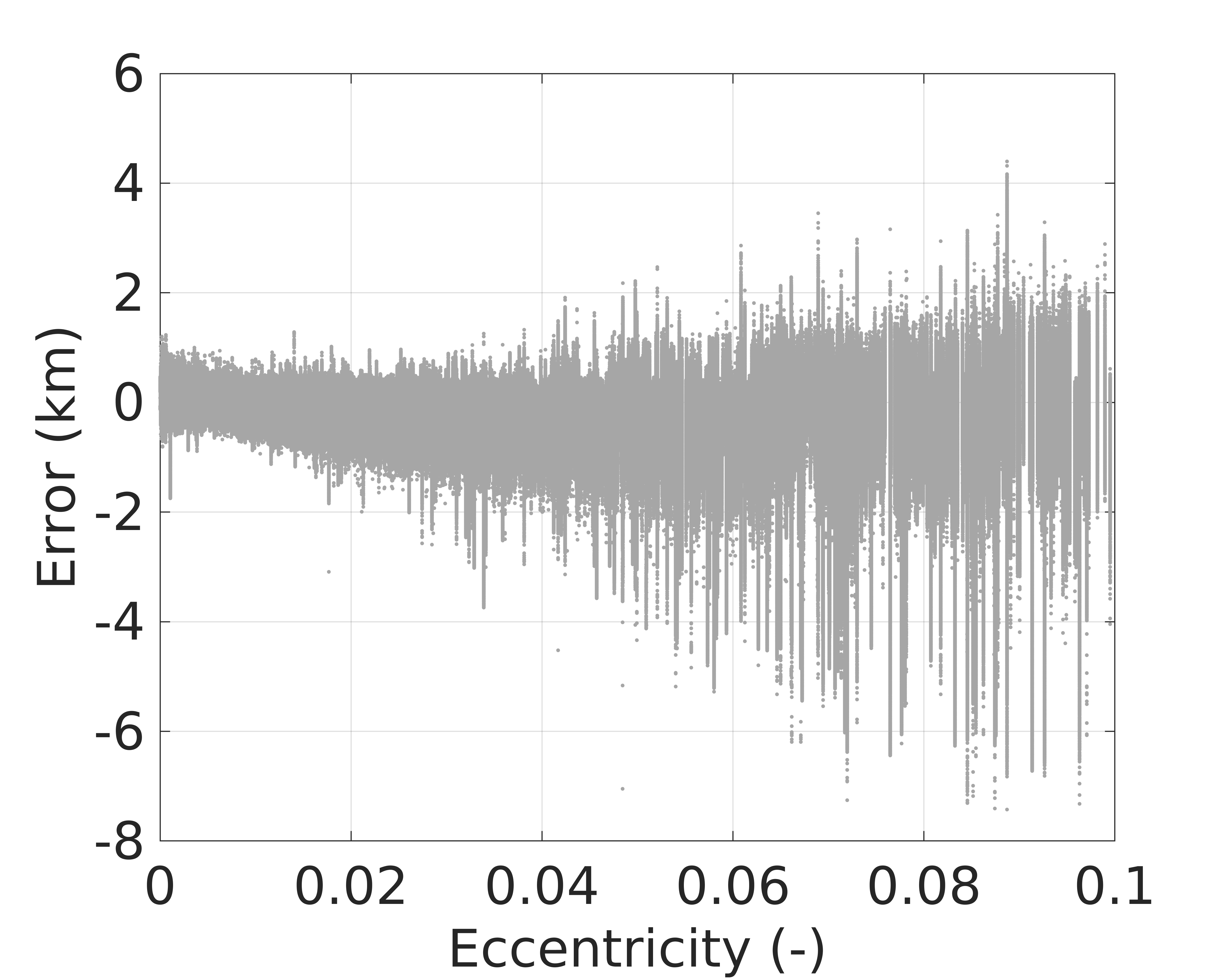}
  \caption{Same errors used to obtain Fig.~\ref{fig:GE_hist} versus
    the eccentricity $\hat{e}$.}
  \label{fig:GE_ECC} 
\end{figure}

Before describing the results of the SOP-filter, the accuracy in the
computation of the bounds of the orbit radius (see
Section~\ref{sec:rbounds}) is assessed. For each of the 30,836,816
pairs, the values of $r_{{\rm min},k}$, $r_{{\rm max},k}$, $r^*_{{\rm
    min},k}$, $r^*_{{\rm max},k}$ ($k=1,2$) obtained from the
analytical procedure are compared to those obtained from numerical
propagations with the high-fidelity dynamical model. In particular, a
positive error means that the estimated maximum/minimum orbit radius
is smaller/larger than the numerical value.  Fig.~\ref{fig:GE_hist}
illustrates the distribution and cumulative density function of the
set made by all these errors (eight values for each pair), revealing a
mean error margin close to 0. Notably, 99\% of the examined cases
displayed an error smaller than 0.39 km, underscoring the robustness
of the algorithm. Moreover, the maximum error is about 4.40
km. Additionally, Fig.~\ref{fig:GE_ECC} displays the same set of
errors of Fig.~\ref{fig:GE_hist} against the eccentricity. The larger
is this orbital element, the higher are the errors in the minimum and
maximum values of the orbit radii.

The performance of the SOP-filter is assessed through the metrics
introduced in Section~\ref{sec:Def_metr}. Next, the results obtained
without applying any buffer are presented in
Sections~\ref{sec:without}. Finally, the analysis of the angular and
radial buffers and the results obtained by applying these buffers are
presented in Sections~\ref{sec:with1} and~\ref{sec:with2}.

\begin{table}
  \centering
  \begin{tabular}{lr}
    \hline
    input orbits & 16,951\\
    almost coplanar pairs & 1,637,190\\
    input pairs & 30,836,816\\
    positive pairs & 6,452,412\\
    \hline
  \end{tabular}
  \caption{SOP-filter input orbits and pairs. The pairs that could
    potentially generate a collision are denoted as ``positive
    pairs''.}
  \label{tab:real_res}
\end{table}
 
\subsection{Performance metrics}
\label{sec:Def_metr}

The following two definitions are relevant to evaluate the performance
of the SOP-filter.

\noindent
A \textit{false positive} occurs when the filter evaluates a pair as
positive because condition~\eqref{eq:pos} is satisfied, while the pair
is flagged as negative from the result of the numerical propagation. A
\textit{false negative} occurs when the filter evaluates a pair as
negative because condition~\eqref{eq:neg} is satisfied, while the pair
is flagged as positive from the result of the numerical
propagation. Usually, one wants to minimize or possibly reduce to zero
the number of false negatives, without increasing too much the number
of false positives.

In order to evaluate the amount of false positives and false negatives
three different metrics are employed, following \cite{SO_filter}.

The \textit{false positives to true positives} ratio is defined as
\begin{equation}
  \rho_{\rm fp} = \frac{N_{\rm fp}}{N_{\rm tp}},
  \label{eq:rhofp}
\end{equation}
where $N_{\rm fp}$ and $N_{\rm tp}$ are the numbers of false and true
positives, respectively.

\noindent
The \textit{false negatives to true positives} ratio is defined as
\[
\rho_{\rm fn} = \frac{N_{\rm fn}}{N_{\rm tp}},
\]
where $N_{\rm fn}$ is the number of false negatives. 

\noindent
Finally, the \textit{filter effectiveness} is
\begin{equation}
  \eta = \frac{N_{\rm out}}{N},
  \label{eq:eta}
\end{equation}
where $N_{\rm out}$ is the number of pairs eliminated by the filter
because they are flagged as negative and therefore can not generate a
collision, and $N$ is the total number of pairs over which the filter
is tested, that is 30,836,816.

\subsection{SOP-filter results without buffers}
\label{sec:without}

Consider first the case where the angular and radial buffers are not
applied. This means that for any pair the intervals $\mathcal{T}_k$,
$\mathcal{T}^*_k$ and $\mathcal{R}_k$, $\mathcal{R}^*_k$ are computed
as described in Sections~\ref{sec:opdom} and
~\ref{sec:rbounds},~\ref{sec:filtering}, respectively, and they are
not inflated.

Table~\ref{tab:results_wb} outlines the results of the SOP-filter,
including the number of false positives and false negatives, along
with the three previously defined metrics. The filter eliminates about
$80.35$\% of the total number of pairs. However, there are too many
false negatives, amounting to about $6.82$\% of the real number of
positive pairs reported in Table~\ref{tab:real_res}. This serious
issue is addressed here by introducing the angular and radial buffers.

\begin{table}
  \centering 
  \begin{tabular}{lr}
    \hline 
    false positives & 46,609 \\
    false negatives & 439,873 \\
    $\rho_{\rm fp}$ & 0.775 \% \\
    $\rho_{\rm fn}$ & 7.316 \% \\
    $\eta$ & 80.351 \% \\
    \hline
  \end{tabular}
  \caption{SOP-filter performance without buffers.}
  \label{tab:results_wb}
\end{table}

\subsection{Buffers analysis}
\label{sec:with1}

Errors affecting the accuracy of the SOP-filter arise mainly from the
location of the mutual nodes and the computation of the orbit
radius. To mitigate these errors and avoid false negatives, the
intervals $\mathcal{T}_k$, $\mathcal{T}^*_k$ and $\mathcal{R}_k$,
$\mathcal{R}^*_k$ are enlarged by means of an angular and radial
buffer, respectively. This approach, while is effective in
eliminating false negatives, might cause a considerable increase in
false positives if the chosen buffers become too large.

The two sources of error are analyzed in Sections~\ref{sec:Int_buf}
and~\ref{sec:R_buf}.

\subsubsection{Angular buffer}
\label{sec:Int_buf}

For each pair of objects, the intervals $\mathcal{T}_k$,
$\mathcal{T}^*_k$ ($k=1,2$) are enlarged by subtracting and adding to
the lower and higher endpoints, respectively, an angular shift. This
quantity will be different for the two endpoints, and its computation
is described below.

\begin{figure}
  \centering
  \includegraphics[width=0.49\textwidth]{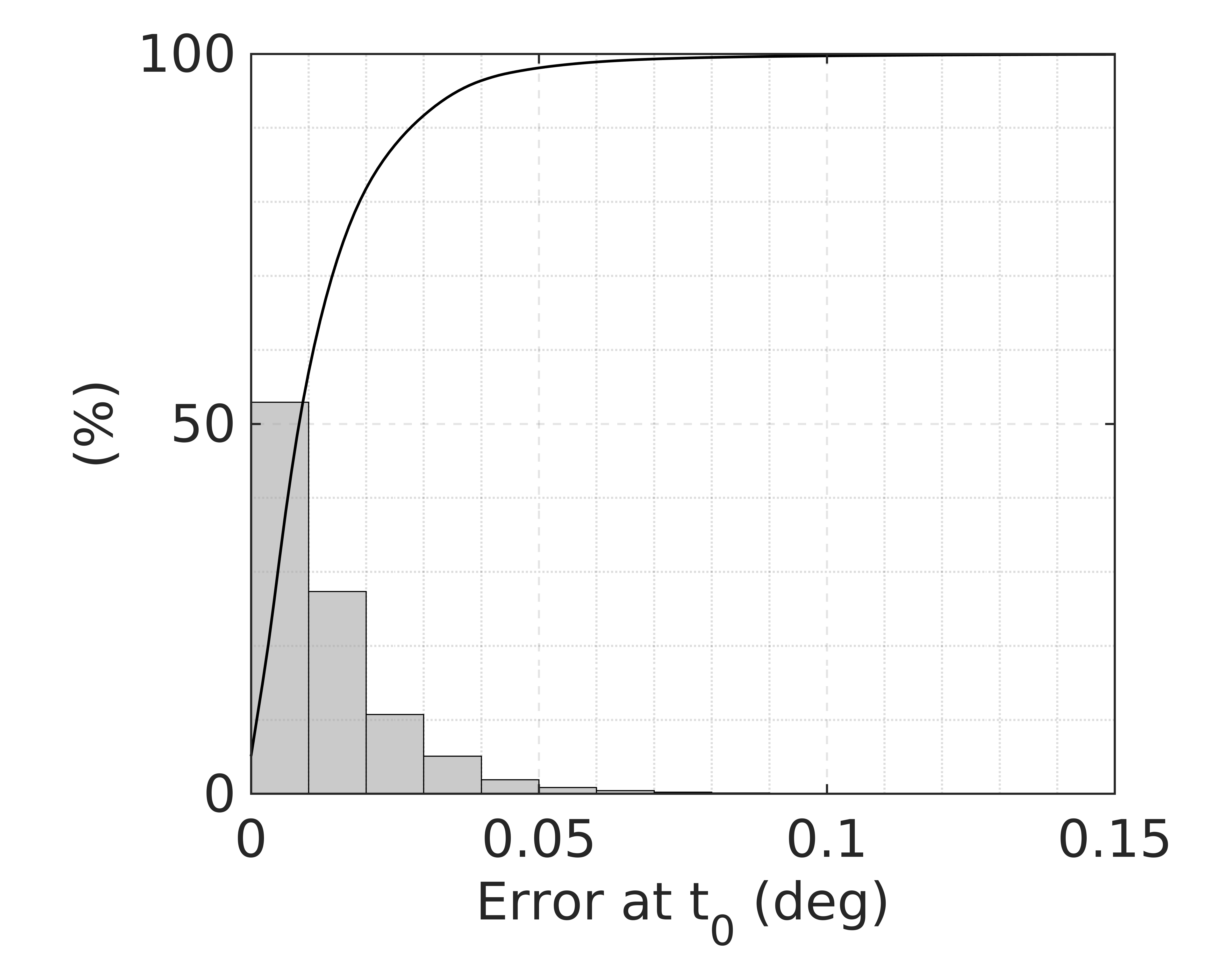}
  \includegraphics[width=0.49\textwidth]{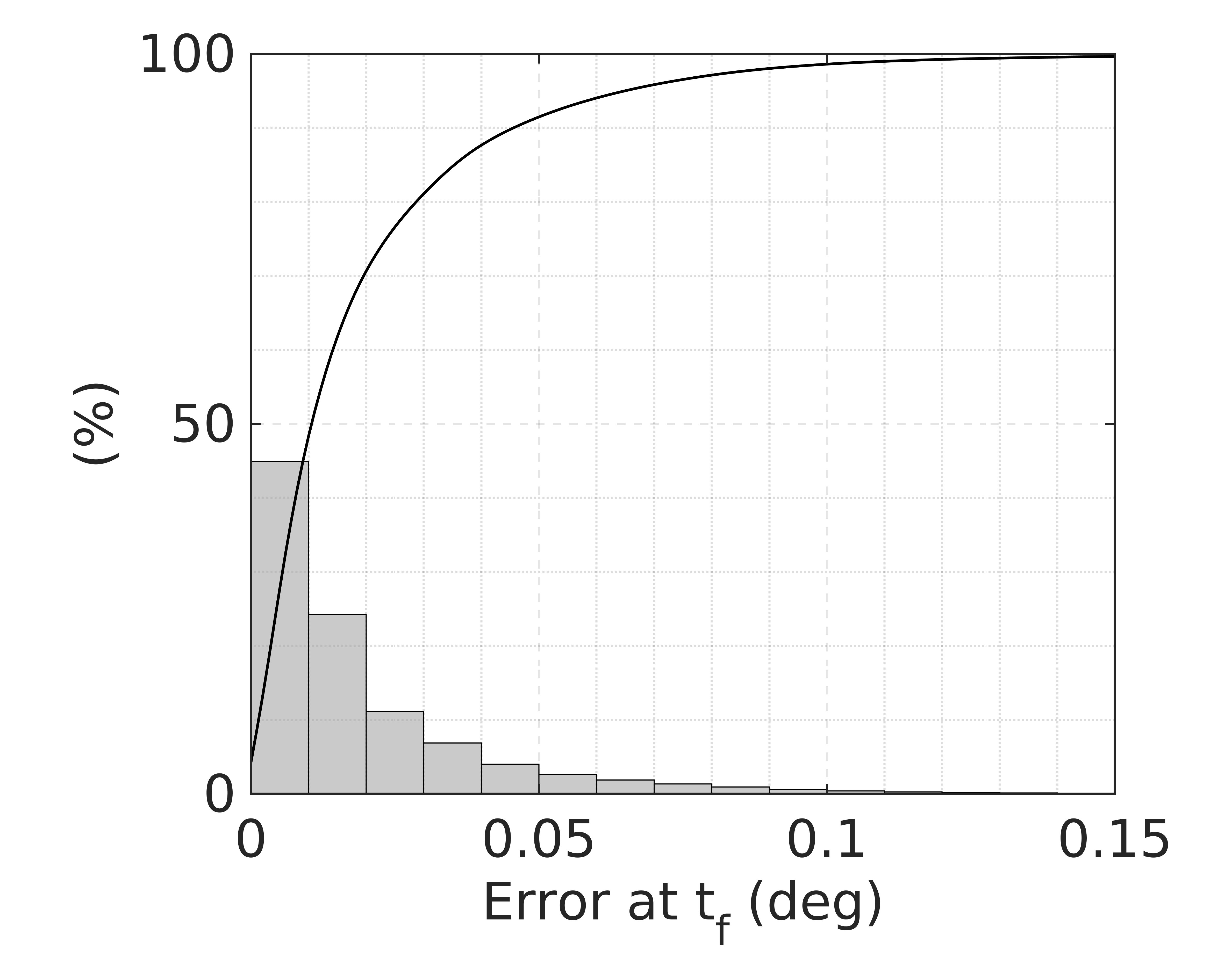}
  \caption{Histogram and cumulative density function of the errors in
    the argument of latitude of the mutual nodes in the Northern
    hemisphere at the initial (left) and final time (right) for the
    input pairs of the SOP-filter.}
  \label{fig:int_error} 
\end{figure}

Consider the $i$-th pair, with $i\in[N]=\{1,\ldots,N\}$. Let
$\hat{\theta}_{k,i}(t_0)$, $\hat{\theta}_{k,i}(t_{\rm f})$ and
$\theta_{k,i}(t_0)$, $\theta_{k,i}(t_{\rm f})$ be the arguments of
latitude of the mutual nodes in the Northern hemisphere at the initial
and final time computed as explained in Section~\ref{sec:opdom} (i.e.,
$\hat{\theta}_k(t_0)$, $\hat{\theta}_k(t_{\rm f})$) and by numerical
propagations with the high-fidelity model, respectively. The
following errors can be defined:
\begin{equation}
  \begin{aligned}
    \epsilon_{k,i}(t_0) &= |\hat{\theta}_{k,i}(t_0)-\theta_{k,i}(t_0)|,\\
    \epsilon_{k,i}(t_{\rm f}) &= |\hat{\theta}_{k,i}(t_{\rm f})-\theta_{k,i}(t_{\rm f})|,
  \end{aligned}
  \label{eq:epski0f}
\end{equation}
with $k=1,2$, $i=1,\ldots,N$. Fig.~\ref{fig:int_error} shows the
distribution and cumulative density function of all the errors
$\epsilon_{k,i}(t_0)$ (left plot) and $\epsilon_{k,i}(t_0)$ (right
plot). Although the mean error in both plots is nearly 0, the errors
at $t_{\rm f}$ display a greater dispersion as expected because of the
approximate analytical prediction of $\hat{\Omega}_k(t_{\rm f})$. At
$t_0$, 99\% of all errors are smaller than 0.062 deg, while at $t_{\rm
  f}$, this value increases to 0.111 deg.

\begin{figure}[h]
  \centering
  \includegraphics[width=0.6\textwidth]{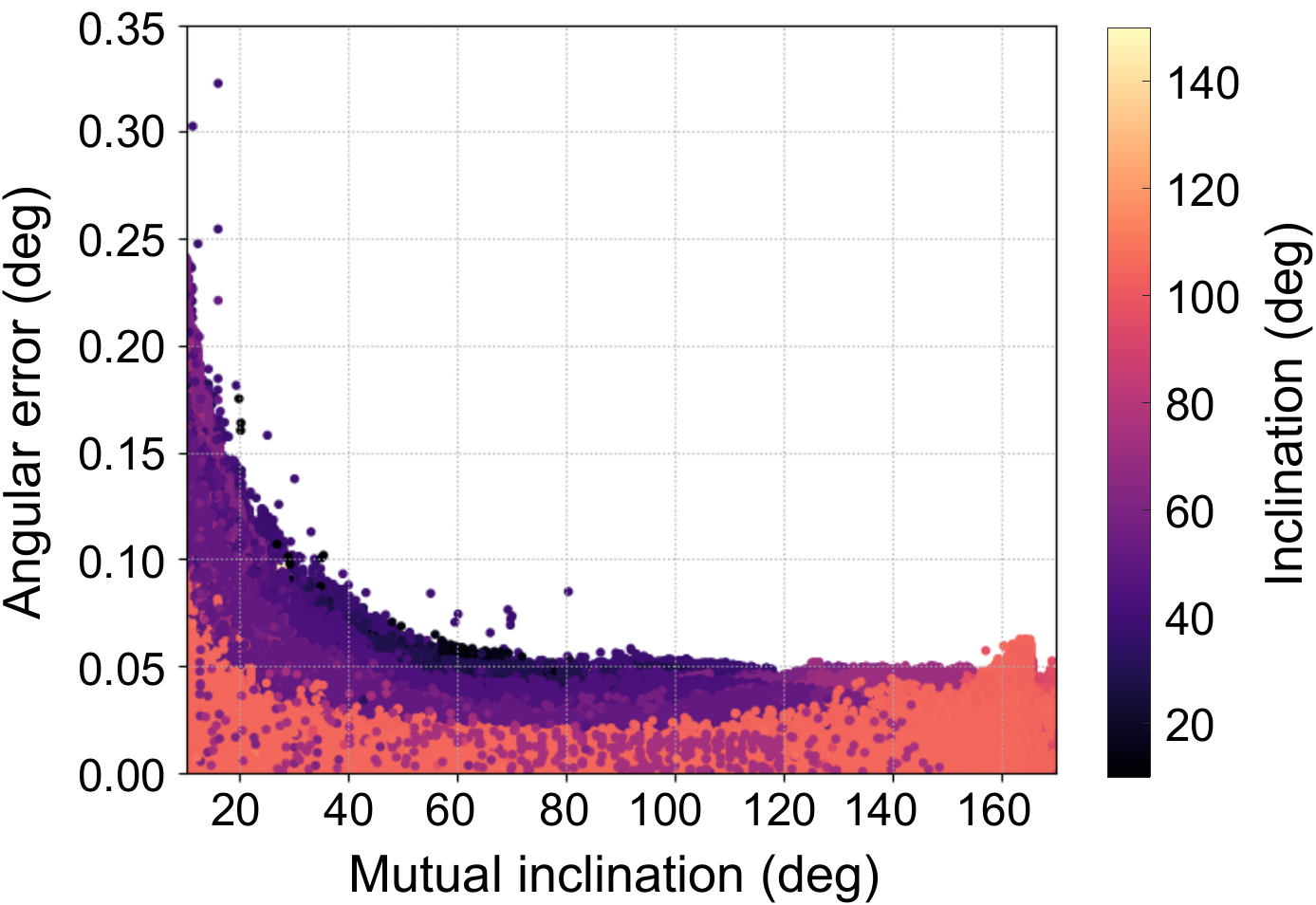}
  \caption{Errors in the argument of latitude of the mutual nodes in
    the Northern hemisphere at the initial time $t_0$ versus the
    mutual inclination $\gamma$ for the input pairs of the
    SOP-filter. The color bar refers to the orbit inclination
    $\hat{i}$.}
  \label{fig:int_mut0} 
\end{figure}

\begin{figure}
  \centering
  \includegraphics[width=0.6\textwidth]{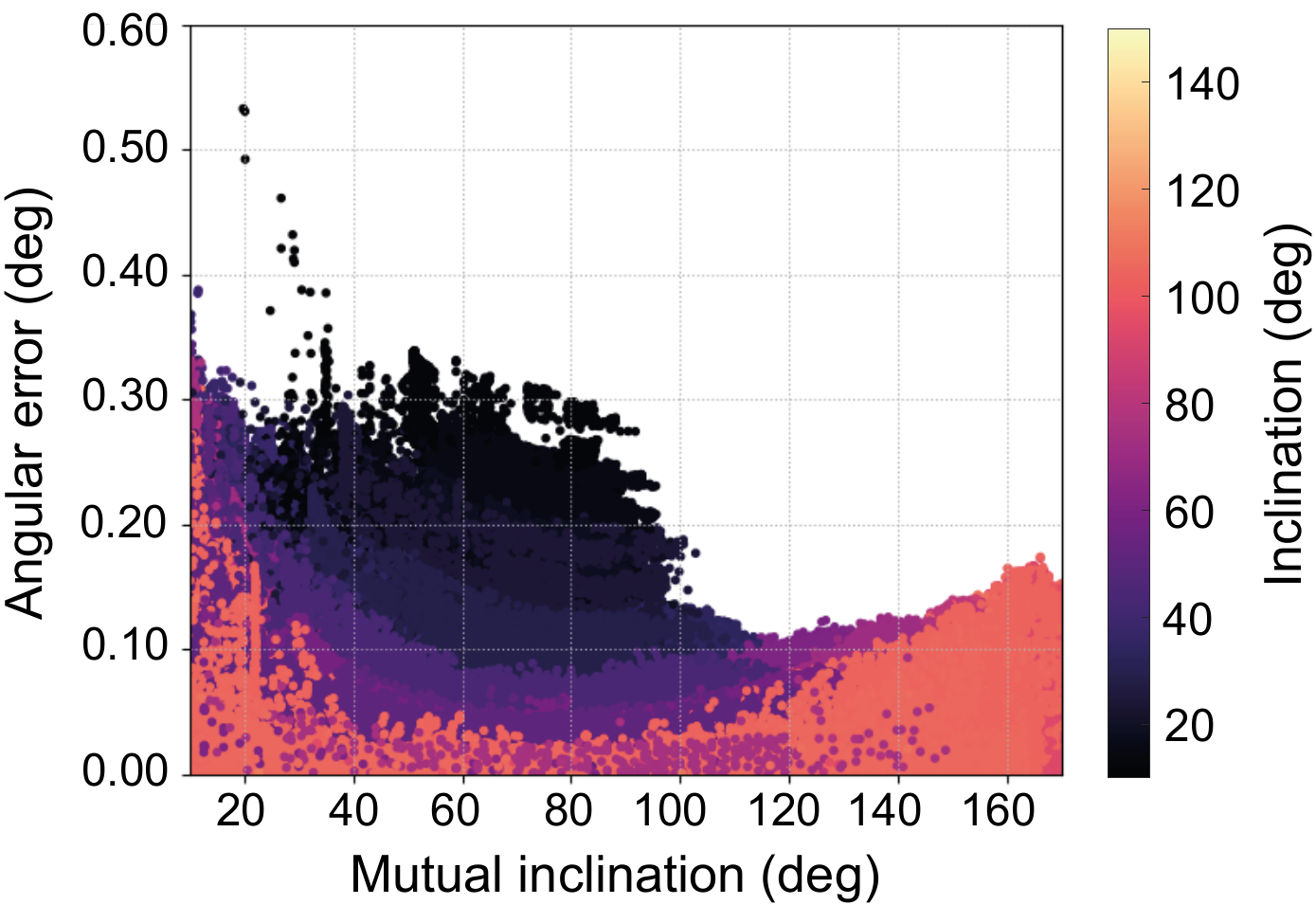}
  \caption{Same as in Fig.~\ref{fig:int_mut0} but at the final time
    $t_{\rm f}$.}
  \label{fig:int_mutf} 
\end{figure}

Figs.~\ref{fig:int_mut0} and~\ref{fig:int_mutf} represent the same
errors of Fig.~\ref{fig:int_error} (left) and Fig.~\ref{fig:int_error}
(right), respectively, versus the mutual inclination. In both figures,
the color bar gives information on the orbit inclination. One can
observe how the error increases as the mutual inclination approaches
10 deg and (even if less appreciably) 170 deg. Moreover,
Fig.~\ref{fig:int_mutf} shows that also small values of the
inclination can lead to a loss of accuracy in the computation of the
arguments of latitude of the mutual nodes at the final epoch $t_{\rm
  f}$.

This analysis reveals that the largest errors can be explained
considering the mutual and orbit inclinations. Therefore, angular
buffers dependent on the values of these quantities ($\gamma$,
$\hat{i}$\,) are introduced as follows. The domain
$\mathcal{S}=[\gamma_0,\pi-\gamma_0]\times[0,\pi)$ of
  $(\gamma,\hat{i}\,)$, where $\gamma_0=\pi/18$, is split into
  $n\times m$ equal subdomains
  \[
  \begin{aligned}
    \mathcal{S}_{(u,v)} = \Bigl\{(\gamma,\hat{i}\,): \gamma_0+\frac{\pi-\gamma_0}{n}(u-1) &
    \le\gamma \le\gamma_0+\frac{\pi-\gamma_0}{n}u,\\[0.3ex]
    \frac{\pi}{m}(v-1)&\le\hat{i}<\frac{\pi}{m}v\Bigr\},
  \end{aligned}
  \]
with $u=1,\ldots,n$, $v=1,\ldots,m$.  Let $\mathcal{F}_{(u,v)}$ be the
subset of $\{1,2\}\times[N]$ containing the pairs of indexes $(k,i)$
corresponding to the object $k$ of the $i$-th pair such that its
mutual and orbit inclinations belong to $\mathcal{S}_{(u,v)}$. Then,
the following angular buffers are introduced for $u=1,\ldots,n$,
$v=1,\ldots,m$:
\begin{align*}
  \delta_0(u,v) &= \max_{(k,i)\in\mathcal{F}_{(u,v)}}\epsilon_{k,i}(t_0),\\
  \delta_{\rm f}(u,v) &= \max_{(k,i)\in\mathcal{F}_{(u,v)}}\epsilon_{k,i}(t_{\rm f}),
\end{align*}
with $\epsilon_{k,i}(t_0)$, $\epsilon_{k,i}(t_{\rm f})$ defined
in~\eqref{eq:epski0f}.  Appendix~\ref{ap:buf} reports the values of
the angular buffers $\delta_0(u,v)$ and $\delta_{\rm f}(u,v)$ for the
different subdomains.

For each object of the input pairs (see Table~\ref{tab:real_res}),
first one determines the values of the indexes $u$, $v$. Then, one
subtracts from (adds to) the lower (higher) endpoints of
$\mathcal{T}_k$ and $\mathcal{T}^*_k$ the angle $\delta_0(u,v)$ if the
considered endpoint was computed from $\hat{\theta}(t_0)$, or the
angle $\delta_{\rm f}(u,v)$ if it was computed either from
$\hat{\theta}(t_{\rm f})$ or from $\hat{i}_1$, $\hat{i}_2$
(see~\ref{eq:thminmax1}). In the latter case, which corresponds to the
situation where the latitude of the mutual nodes in the Northern
hemisphere reaches its maximum value, this buffer is not optimal,
however, only a small percentage of pairs falls in this case.

\subsubsection{Radial buffer}
\label{sec:R_buf}

The expression of the orbit radius employed by the SOP-filter is given
in~\eqref{eq:radius2} as a function of $\beta$ and $\hat{\theta}$,
which are regarded as two independent variables. This assumption is
motivated by the very separated time scales of their evolutions.

The accuracy of equation~\eqref{eq:radius2} is assessed by computing
the values of $r$ for $\hat{\theta}\in[0,2\pi)$ and for
  $\tau\in[\tau_0,\tau_0+T_C]$, where $\tau_0$ is the initial epoch,
  fixed at 2,459,885.89 JD, and
  $T_C=2\pi/\sqrt{\kappa_\xi\kappa_\eta}$ corresponds to one period of
  revolution of the apsidal line according to R. Cook's extended
  theory. Let $\beta_0$, $\beta_C=\beta_0+{\rm sgn}\,\kappa_\xi 2\pi$
  be the values of $\beta$ obtained from equation~\eqref{eq:beta} with
  $\tau=\tau_0,\tau_0+T_C$, respectively. The domain
  $[0,2\pi)\times[\beta_0,\beta_C]$ for the variables
    $(\hat{\theta},\beta)$ is sampled as follows
\[
(\hat{\theta}_j,\beta_s) = \left(\frac{2\pi}{m}(j-1),\beta_0+{\rm sgn}\,\kappa_\xi\frac{2\pi}{m}s\right),
\qquad (j,s)\in\mathcal{G},
\]
where $\mathcal{G}=\{(j,s):1\le j\le m,0\le s\le m\}$. Let $\tau_s$ be
the value of $\tau$ corresponding to $\beta_s$ through
equation~\eqref{eq:beta}.

Consider the $n$-th object, with $n\in[N_{\rm ob}]=\{1,\ldots,N_{\rm
  ob}\}$ and $N_{\rm ob}$ is the number of input orbits (see
Table~\ref{tab:real_res}). The following error can be defined:
\begin{equation}
  \epsilon_n=\max_{(j,s)\in\mathcal{G}}|r_n(\hat{\theta}_j,\beta_s)-\mathfrak{r}_n(\hat{\theta}_j;\tau_s)|,
  \label{eq:epsn}
\end{equation}
where $r_n(\hat{\theta},\beta)$ is the orbit radius computed
from~\eqref{eq:radius2} and
\[
\mathfrak{r}_n(\hat{\theta};\tau)=\frac{a_n(\tau)(1-e_n(\tau)^2)}{1+e_n(\tau)\cos(\hat{\theta}-\omega_n(\tau))},
\]
with the osculating orbital elements $a_n$, $e_n$, and $\omega_n$ of
the $n$-th object obtained by numerical propagation.

\begin{figure}[h!]
  \centering
  \includegraphics[width=0.49\textwidth]{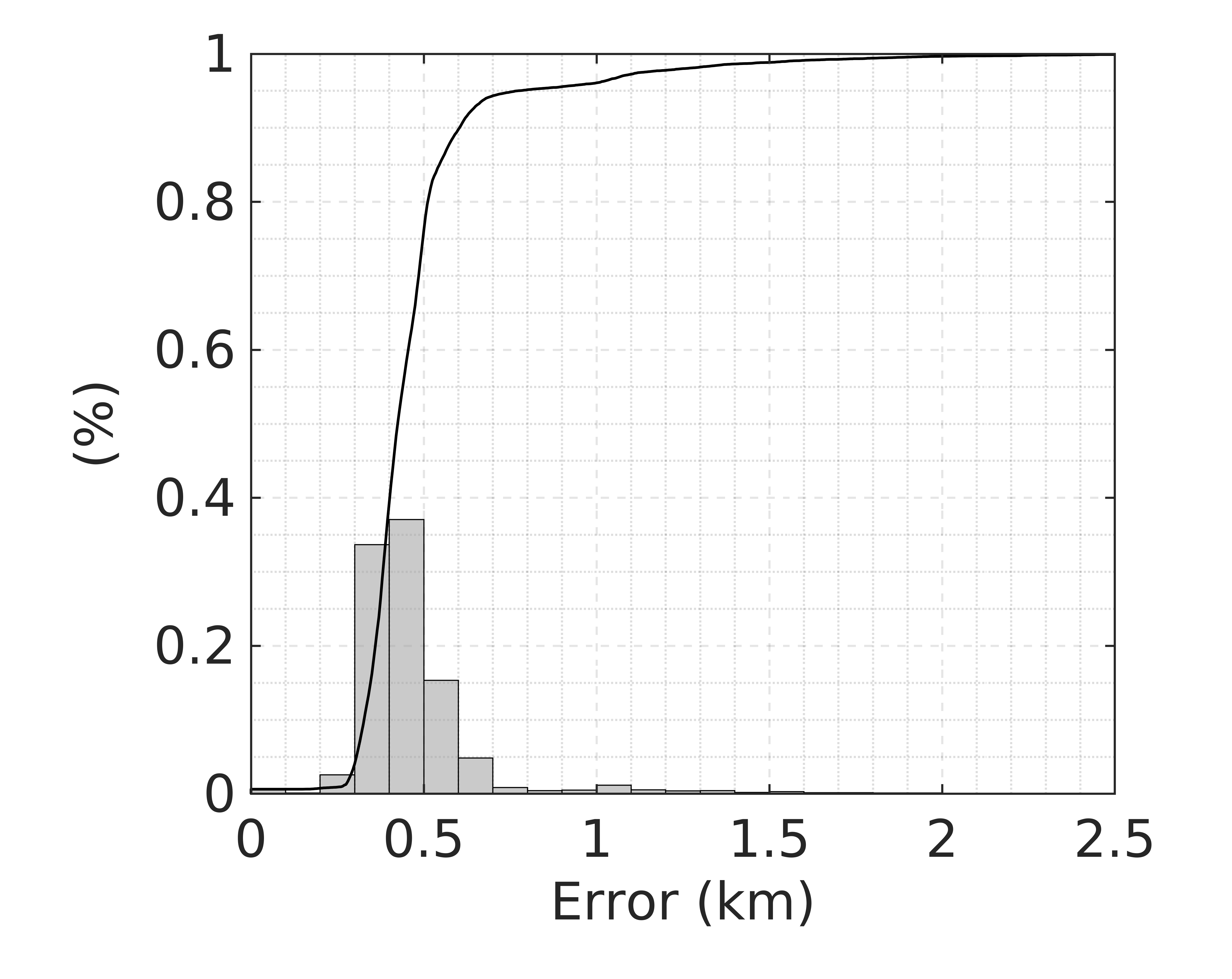}
  \caption{Histogram and cumulative density function of the radial
    error $\epsilon_n$ defined in~\eqref{eq:epsn} for the input
    objects of the SOP-filter.}
  \label{fig:r_error} 
\end{figure}

Fig.~\ref{fig:r_error} depicts the distribution and cumulative density
function of all the errors $\epsilon_n$. The mean value of the
distribution is about 0.5 km, and for 99\% of the input objects the
radial error is smaller than 1.55 km. On the other hand, the maximum
error is much larger, amounting to about 4.88 km. Setting the radial
buffer equal to this value would be too conservative for most of the
orbits, and thus is far from being an optimal choice. A better
approach is to adapt the size of the buffer based on those features of
the orbits that impact the error.

\begin{figure}
  \centering
  \includegraphics[width=0.6\textwidth]{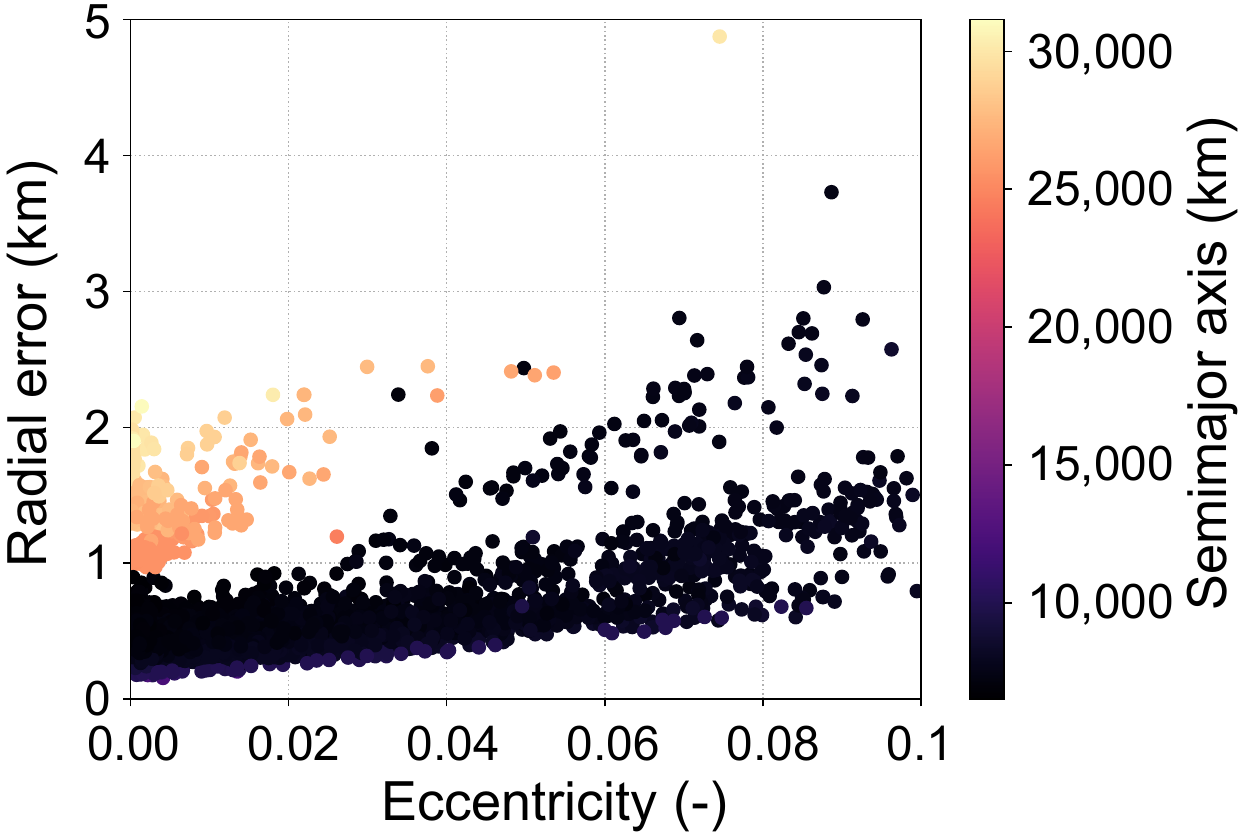}
  \caption{Radial error $\epsilon_n$ defined in~\eqref{eq:epsn} for
    the input objects of the SOP-filter versus the eccentricity
    $\hat{e}$. The color bar refers to the semi-major axis
    $\hat{a}$.}
  \label{fig:r_a} 
\end{figure}

\begin{figure}[h]
  \centering
  \includegraphics[width=0.6\textwidth]{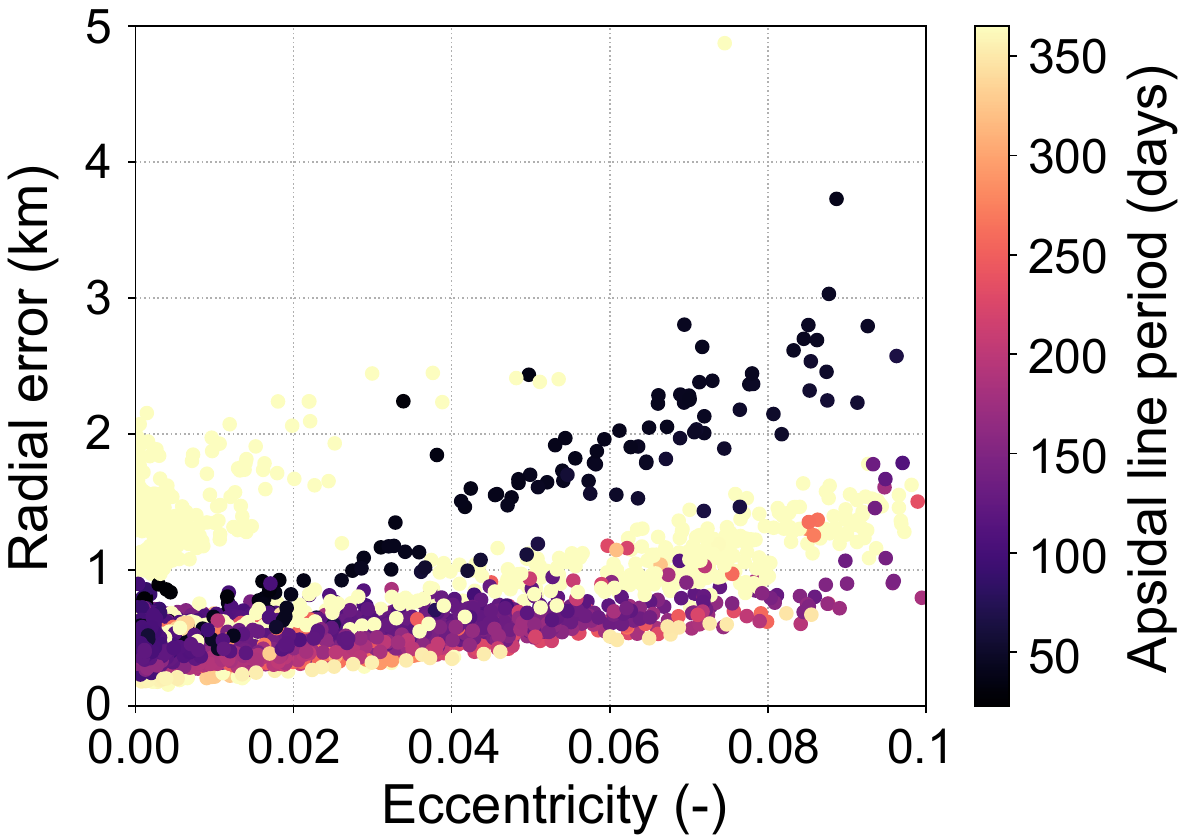}
  \caption{Same as in Fig.~\ref{fig:r_a}, but the color bar refers to
    the precession period of the apsidal line given by R. Cook's
    extended theory.}
  \label{fig:r_Cp} 
\end{figure}

It was shown in Fig.~\ref{fig:GE_ECC} that the eccentricity influences
the error in the orbit radius. Here, Figs.~\ref{fig:r_a}
and~\ref{fig:r_Cp} display the radial errors $\epsilon_n$ versus the
eccentricity. The color bars denote the semi-major axis
(Fig.~\ref{fig:r_a}) and the precession period $T_C$ of the line of
apsides (Fig.~\ref{fig:r_Cp}). From Fig.~\ref{fig:r_a} it is evident
that higher altitudes lead to larger errors. Indeed,
equation~\eqref{eq:radius2} does not take into account third-body
effects, which become more relevant as the semi-major axis
increases. Notably, all orbits with eccentricities smaller than 0.02
and a radial error exceeding 1 km have a semi-major axis greater than
10,000 km. Concerning the influence of $T_C$, one can observe from
Fig.~\ref{fig:r_Cp} that for shorter precession periods, the increase
of the error with the eccentricity is more pronounced. This is
expected, since a smaller $T_C$ implies a faster change in $\beta$,
which instead is assumed fixed while $\hat{\theta}$ is varied in
$[0,2\pi)$.

In light of these results, the value of the radial buffer will depend
on the semi-major axis ($\hat{a}$), eccentricity ($\hat{e}$), and
period $T_C$. Following a similar procedure as in the previous
section, the space $\hat{a}>0$, $\hat{e}\in[0,0.1]$, $T_C>0$ is split
into a finite number of subdomains. For each of them, the maximum of
the errors $\epsilon_n$ (defined in equation~\ref{eq:epsn}) is
computed over the indexes $n$ of those objects whose mean orbital
elements $\hat{a}$, $\hat{e}$, and period $T_C$ belong to the
considered subdomain. The resulting value defines the radial buffer of
that subdomain.

For the $k$-th object ($k=1,2$) of the $i$-th pair ($i=1,\ldots,N$),
the corresponding radial buffer is subtracted from (added to) the
lower (higher) endpoints of both the radial intervals $\mathcal{R}_k$,
$\mathcal{R}_k^*$.  Appendix~\ref{ap:buf} reports the radial buffers
of each subdomain.

\subsection{SOP-filter results with buffers}
\label{sec:with2}

The results of the SOP-filter with both the interval and radial
buffers are presented.  The former enlarge the domains
$\mathcal{D}_k$, $\mathcal{D}^*_k$ where the minimum and maximum of
the orbit radius of the object $k$ of a given pair are searched
for. The latter enlarges the intervals $\mathcal{R}_k$,
$\mathcal{R}^*_k$ that are employed to decide whether a pair is
positive or negative (see Section~\ref{sec:filtering}).

Table~\ref{results_wb} summarises the outcomes obtained by applying
the SOP-filter with the angular and radial buffers as detailed in
Section~\ref{sec:with1}. The filter excludes 75\% of the input pairs
from further conjunction assessment. Moreover, the number of pairs is
reduced from 143,659,725 to 9,313,513 by the SO- and SOP-filters.

\begin{table}
  \centering 
  \begin{tabular}{lr} 
    \hline 
    false positives & 1,223,911 \\
    false negatives & 0 \\
    $\rho_{\rm fp}$ & 18.968 \% \\
    $\eta$ & 75.107 \%\\
    \hline
  \end{tabular}
  \caption{SOP-filter performance with buffers.}
  \label{results_wb}
\end{table}

In conjunction analysis a safety distance $D$ is commonly introduced
to take into account of the uncertainty in the orbital state: a
conjunction occurs if the distance between two objects is smaller
than $D$.  Different choices of $D$ can be found in the literature
\citep[see for example][]{Khutorovsky93,Healy,Wood,
  Alfano12,Casanova}.

\begin{figure}[h]
  \centering
  \includegraphics[width=\textwidth]{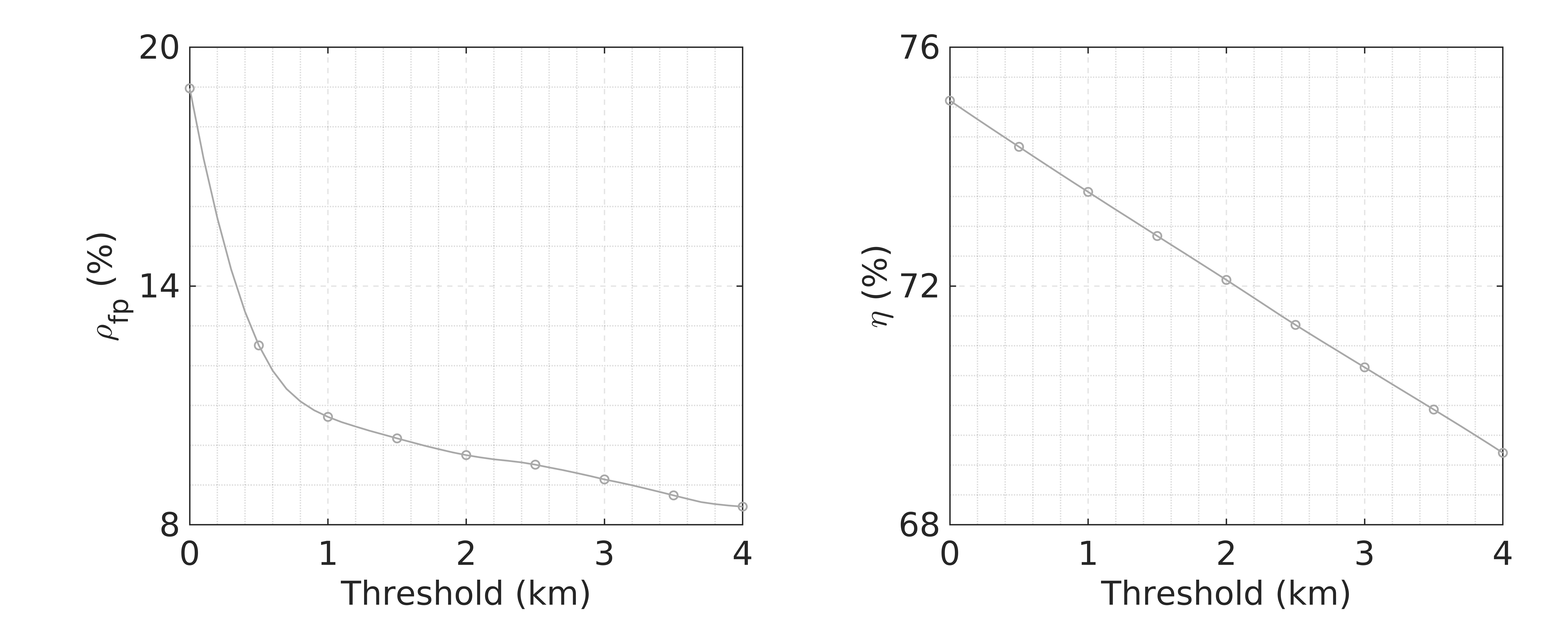}
  \caption{False positives to true positive ratio (left,
    see~\ref{eq:rhofp}) and filter effectiveness (right,
    see~\ref{eq:eta}) versus the applied threshold distance $D$.}
  \label{fig:thres} 
\end{figure}

A threshold distance $D$ is considered also in this work to further
enlarge the intervals $\mathcal{R}_k$,
$\mathcal{R}^*_k$. Fig.~\ref{fig:thres} shows how the SOP-filter's
performance metrics $\rho_{\rm fp}$ and $\eta$ decrease as the
distance $D$ is increased from 0 to 4 km.

Comparing the proposed SOP-filter with similar methods existing in the
literature is not an easy task. Here, a comparison is made with the
filter sequence implemented by \citet{Casanova}, where an \emph{all
  vs. all} conjunction analysis of 864 objects is carried out. Their
orbit path Filters I and II excluded 323,077 pairs, representing
86.7\% of the total number of pairs. Notably, Filter I alone
eliminated 41.1\% of pairs, while Filter II further eliminated 77.2\%
of the remaining pairs. Employing the same threshold $D=100$~m as in
\citet{Casanova}, the SOP-filter removed 74.95\% of the pairs received
from the SO-filter, which was able to exclude 77.40\% of the input
pairs. Therefore, the application of the SO- and SOP-filters reduced
of 93.5\% the number of pairs to be investigated, highlighting a
considerable improvement.

Finally, computations were performed on a machine with an AMD EPYC
7542 32-Core Processor, with a base frequency of 1.5 GHz and max
frequency of 2.9 GHz. The code was implemented in MATLAB R2021b,
utilizing the Parallel Computing Toolbox with 13 cores. With these
specifications, the computational time required by the SOP-filter to
process the 30,836,816 pairs was 4.07 hours.

\section{Conclusions}
\label{sec:conclusions}

The work introduces a novel approach for the estimation of the
distance between perturbed orbital paths and for the implementation of
an effective path filter to treat perturbed orbits in the Earth
environment.  The effect of environmental perturbations dominated by
zonal harmonics is shown to play a key role in determining the true
distance between orbital trajectories which makes the use of classical
Keplerian MOID computation schemes unsuitable.  In order to correctly
incorporate these perturbations and treat orbits of small to moderate
eccentricity (hence covering the great majority of resident space
objects orbiting the Earth) an extension of the linear secular theory
proposed by Cook (1966) is employed, which is an additional
contribution of this work. An analytical procedure to estimate the
upper and lower bound of radial distances at nodal crossings is
derived accounting for short-period variations of the semi-major axis,
eccentricity, and mean anomaly.  In order to safely apply these
analytical results to conjunction filtering and reduce false negatives
to ideally zero, an angular and a radial buffer are implemented and
carefully evaluated and tailored to the specific pair, thus avoiding
an excessive growth of false positives.  Extensive numerical tests
show that the proposed filter is highly efficient, also in terms of
computational time. After applying the filter to a dataset of 16,951
resident space objects, it eliminates 75.11\% of the total pairs with
no false negatives detected.  Finally, when applied together with a
previously developed space occupancy filter \citep[see][]{SO_filter}
92.3\% of the initial pairs are eliminated from further analysis. This
dramatic reduction in the number of pairs has significant implications
for space traffic management and collision risk mitigation.

\section*{Acknowledgements}

A.~S. Rivero and R. Vazquez gratefully acknowledge support by grant
TED2021-132099B-C33 funded by MICIU/AEI/10.13039/501100011033 and by
``European Union NextGenerationEU/PRTR''. Additionally, A.~S. Rivero
was funded by a FPU grant from the Spanish Ministry of
Universities. G. Ba\`u carried out this study within the Space It Up
project funded by the Italian Space Agency, ASI, and the Ministry of
University and Research, MUR, under contract n. 2024-5-E.0 - CUP
n. I53D24000060005.



\bibliographystyle{plainnat}
\bibliography{references} 

\begin{thebibliography}{28}
\providecommand{\natexlab}[1]{#1}
\providecommand{\url}[1]{\texttt{#1}}
\expandafter\ifx\csname urlstyle\endcsname\relax
  \providecommand{\doi}[1]{doi: #1}\else
  \providecommand{\doi}{doi: \begingroup \urlstyle{rm}\Url}\fi

\bibitem[Alarc{\'o}n~Rodr{\'\i}guez et~al.(2002)Alarc{\'o}n~Rodr{\'\i}guez,
  Mart{\'\i}nez~Fadrique, and Klinkrad]{smartsieve}
J.~R. Alarc{\'o}n~Rodr{\'\i}guez, F.~M. Mart{\'\i}nez~Fadrique, and
  H.~Klinkrad.
\newblock Collision risk assessment with a ``smart sieve'' method.
\newblock \emph{European Space Agency, (Special Publication) ESA SP},
  486:\penalty0 159--164, 2002.

\bibitem[Alfano(2012)]{Alfano12}
S.~Alfano.
\newblock Toroidal path filter for orbital conjunction screening.
\newblock \emph{Celestial Mechanics and Dynamical Astronomy}, 113\penalty0
  (3):\penalty0 321--334, 2012.

\bibitem[Battin(1999)]{battin}
R.~H. Battin.
\newblock \emph{An Introduction to the Mathematics and Methods of
  Astrodynamics}.
\newblock AIAA, Inc., 1999.
\newblock p. 504.

\bibitem[Bonaccorsi et~al.(2024)Bonaccorsi, Montaruli, Di~Lizia, Peroni,
  Panico, Rigamonti, and Del~Prete]{bonaccorsi2023software}
S.~Bonaccorsi, M.~F. Montaruli, P.~Di~Lizia, M.~Peroni, A.~Panico,
  M.~Rigamonti, and F.~Del~Prete.
\newblock Conjunction analysis software suite for space surveillance and
  tracking.
\newblock \emph{Aerospace}, 11\penalty0 (2), 2024.

\bibitem[Casanova et~al.(2014)Casanova, Tardioli, and Lema{\^{i}}tre]{Casanova}
D.~Casanova, C.~Tardioli, and A.~Lema{\^{i}}tre.
\newblock Space debris collision avoidance using a three-filter sequence.
\newblock \emph{MNRAS}, 442\penalty0 (4):\penalty0 3235--3242, 2014.

\bibitem[Cook(1966)]{cook1966perturbations}
G.~E. Cook.
\newblock Perturbations of near-circular orbits by the earth's gravitational
  potential.
\newblock \emph{Planetary and Space Science}, 14\penalty0 (5):\penalty0
  433--444, 1966.

\bibitem[Cook(1992)]{cook1992}
R.~A. Cook.
\newblock The long-term behavior of near-circular orbits in a zonal gravity
  field.
\newblock \emph{Advances in the Astronautical Sciences}, 76:\penalty0
  2205--2221, 1992.

\bibitem[Cox et~al.(1992)Cox, Little, and O'Shea]{Cox}
D.~Cox, J.~Little, and D.~O'Shea.
\newblock \emph{Ideals, Varieties, and Algorithms}.
\newblock Springer-Verlag, 1992.

\bibitem[Escobar et~al.(2012)Escobar, {\'A}gueda, Mart{\'i}n, and
  Mart{\'i}nez]{Escobar}
D.~Escobar, A.~{\'A}gueda, L.~Mart{\'i}n, and F.~M. Mart{\'i}nez.
\newblock Efficient all vs. all collision risk analyses.
\newblock \emph{Journal of Aerospace Engineering, Sciences and Applications},
  4\penalty0 (2):\penalty0 40--48, 2012.

\bibitem[Gronchi(2003)]{Gronchi02}
G.~F. Gronchi.
\newblock On the stationary points of the squared distance between two ellipses
  with a common focus.
\newblock \emph{SIAM Journal on Scientific Computing}, 24\penalty0
  (1):\penalty0 61--80, 2003.

\bibitem[Gronchi(2005)]{Gronchi05}
G.~F. Gronchi.
\newblock An algebraic method to compute the critical points of the distance
  function between two keplerian orbits.
\newblock \emph{Celestial Mechanics and Dynamical Astronomy}, 93:\penalty0
  295--329, 2005.

\bibitem[Gronchi et~al.(2023)Gronchi, Ba{\`u}, and
  Grassi]{gronchi2023revisiting}
G.~F. Gronchi, G.~Ba{\`u}, and C.~Grassi.
\newblock Revisiting the computation of the critical points of the keplerian
  distance.
\newblock \emph{Celestial Mechanics and Dynamical Astronomy}, 135\penalty0
  (5):\penalty0 48, 2023.

\bibitem[Healy(1995)]{Healy}
L.~M. Healy.
\newblock Close conjunction detection on parallel computer.
\newblock \emph{Journal of Guidance, Control, and Dynamics}, 18\penalty0
  (4):\penalty0 824--829, 1995.

\bibitem[Hedo et~al.(2018)Hedo, Ru{\'\i}z, and Pel{\'a}ez]{hedo2018minimum}
J.~M. Hedo, M.~Ru{\'\i}z, and J.~Pel{\'a}ez.
\newblock On the minimum orbital intersection distance computation: a new
  effective method.
\newblock \emph{MNRAS}, 479\penalty0 (3):\penalty0 3288--3299, 2018.

\bibitem[Hedo et~al.(2020)Hedo, Fantino, Ru{\'\i}z, and
  Pel{\'a}ez]{hedo2020minimum}
J.~M. Hedo, E.~Fantino, M.~Ru{\'\i}z, and J.~Pel{\'a}ez.
\newblock Minimum orbital intersection distance: an asymptotic approach.
\newblock \emph{A\&A}, 633:\penalty0 A22, 2020.

\bibitem[Hoots et~al.(1984)Hoots, Crawford, and Roehrich]{Hoots84}
F.~R. Hoots, L.~L. Crawford, and R.~L. Roehrich.
\newblock An analytic method to determine future close approaches between
  satellites.
\newblock \emph{Celestial Mechanics}, 33\penalty0 (2):\penalty0 143--158, 1984.

\bibitem[Kaula(1966)]{Kaula}
W.~M. Kaula.
\newblock \emph{Theory of Satellite Geodesy: Applications of Satellites to
  Geodesy}.
\newblock Blaisdell Publishing Company, 1966.

\bibitem[Kerr and S{\'a}nchez~Ortiz(2021)]{Kerr21}
E.~Kerr and N.~S{\'a}nchez~Ortiz.
\newblock State of the art and future needs in conjunction analysis methods,
  processes and software.
\newblock In \emph{Proceedings of the 8th European Conference on Space Debris
  (virtual), Darmstadt, Germany, 20–23 April 2021}, 2021.

\bibitem[Khutorovsky et~al.(1993)Khutorovsky, Boikov, and
  Kamensky]{Khutorovsky93}
Z.~N. Khutorovsky, V.~F. Boikov, and S.~Y. Kamensky.
\newblock Direct method for the analysis of collision probability of artificial
  space objects in leo: techniques, results and applications.
\newblock In \emph{Proceedings of the First European Conference on Space
  Debris, Darmstadt, Germany, 5-7 April 1993}, pages 491--508, 1993.

\bibitem[Kozai(1959)]{Kozai}
Y.~Kozai.
\newblock The motion of a close earth satellite.
\newblock \emph{The Astronomical Journal}, 64:\penalty0 367--377, 1959.

\bibitem[Kozai(1962)]{Kozai62}
Y.~Kozai.
\newblock Second-order solution of artificial satellite theory without air
  drag.
\newblock \emph{The Astronomical Journal}, 67\penalty0 (7):\penalty0 446--461,
  1962.

\bibitem[Lyddane(1963)]{Lyddane}
R.~H. Lyddane.
\newblock Small eccentricities or inclinations in the brouwer theory of the
  artificial satellite.
\newblock \emph{The Astronomical Journal}, 68\penalty0 (8):\penalty0 555--558,
  1963.

\bibitem[Muelhaupt et~al.(2019)Muelhaupt, Marlon, Morin, and
  Wilson]{Muelhaupt19}
T.~J. Muelhaupt, E.~S. Marlon, J.~Morin, and R.~S. Wilson.
\newblock Space traffic management in the new space era.
\newblock \emph{Journal of Space Safety Engineering}, 6\penalty0 (2):\penalty0
  80--87, 2019.

\bibitem[Rivero et~al.(2024)Rivero, Bombardelli, and Vazquez]{SO_filter}
Ana~S. Rivero, Claudio Bombardelli, and Rafael Vazquez.
\newblock Short-term space occupancy and conjunction filter, 2024.
\newblock URL \url{https://arxiv.org/abs/2309.02379}.

\bibitem[Smart(1953)]{smart1953celestial}
W.~M Smart.
\newblock \emph{Celestial Mechanics}.
\newblock Longmans, Green \& Co, 1953.

\bibitem[Stevenson et~al.(2023)Stevenson, Rodriguez-Fernandez, Urrutxua, and
  Camacho]{Stevenson2023}
E.~Stevenson, V.~Rodriguez-Fernandez, H.~Urrutxua, and D.~Camacho.
\newblock Benchmarking deep learning approaches for all-vs-all conjunction
  screening.
\newblock \emph{Advances in Space Research}, 72\penalty0 (7):\penalty0
  2660--2675, 2023.

\bibitem[Vallado et~al.(2006)Vallado, Crawford, Hujsak, and
  Kelso]{vallado2006revisiting}
D.~A Vallado, P.~Crawford, R.~Hujsak, and T.~S. Kelso.
\newblock Revisiting spacetrack report\# 3: Rev 1.
\newblock In \emph{AIAA/AAS Astrodynamics Specialist Conference and Exhibit,
  21-24 August 2006, Keystone, CO}, 2006.

\bibitem[Woodburn et~al.(2010)Woodburn, Coppola, and Stoner]{Wood}
J.~Woodburn, V.~Coppola, and F.~Stoner.
\newblock A description of filters for minimizing the time required for orbital
  conjunction computations.
\newblock \emph{Advances in the Astronautical Sciences}, 135:\penalty0
  1157--1173, 2010.

\end{thebibliography}



\appendix

\section{Polynomial coefficients}
\label{ap:coef}

For convenience, let us introduce
\[
\sigma = \sqrt{\frac{\kappa_\eta}{\kappa_\xi}},\quad
K = \frac{J_2}{4\hat{a}},\quad
\kappa = \sin^2\hat{i}.
\]

\noindent
The coefficients of the polynomial $q(x)$, introduced
in~\eqref{eq:pq}, are given by:
\begin{equation*}
  \begin{split}
    q_0 &= \frac{\hat{a}}{4}\Bigl[4e_f(e_f^2-1)-2e_fe_p^2(2-3\sigma^2)+8e_fe_p\cos{\beta_*}\\
      &\quad -2e_fe_p^2(2+3\sigma^2)\cos 2\beta_*-e_p\sigma\bigl(4-12e_f^2-e_p^2(2\\
      &\quad -3\sigma^2)\bigr)\sin\beta_*+4e_p^2\sigma\sin 2\beta_*-e_p^3\sigma(2+\sigma^2)\sin 3\beta_*\Bigr],
  \end{split}
\end{equation*}
\begin{equation*}
  \begin{split}
    q_1 &= 2\hat{a}\bigl[2e_f^2-e_p^2(1-\sigma^2)\bigr]-8K\kappa-\frac{1}{2}\hat{a}e_p\bigl[28e_f^2\\
      &\quad +e_p^2(7\sigma^2-6)-4\bigr]\cos\beta_*-2\hat{a}e_p^2(1+\sigma^2)\cos 2\beta_*\\
      &\quad +\hat{a}e_p^3\Bigl(1+\frac{7}{2}\sigma^2\Bigr)\cos 3\beta_*
      +2\hat{a}e_fe_p\sigma(4\sin\beta_*-7e_p\sin 2\beta_*),
  \end{split}
\end{equation*}
\begin{equation*}
  \begin{split}
    q_2 &= -\frac{\hat{a}}{4}\Bigl[4e_f(1+11e_f^2)-2e_fe_p^2(34-33\sigma^2)+40e_fe_p\cos{\beta_*}\\
      &\quad -2e_fe_p^2(34+33\sigma^2)\cos 2\beta_*+e_p\sigma\bigl(4+132e_f^2-e_p^2(34\\
      &\quad -33\sigma^2)\bigr)\sin\beta_*+20e_p^2\sigma\sin 2\beta_*-e_p^3\sigma(34+11\sigma^2)\sin 3\beta_*\Bigr],
  \end{split}
\end{equation*}
\begin{equation*}
  \begin{split}
    q_3 &= 2\hat{a}e_p\cos\beta_*\Bigl[2+22e_f^2-e_p^2(4-11\sigma^2)\\
      &\quad -e_p^2(4+11\sigma^2)\cos 2\beta_*+44e_fe_p\sigma\sin\beta_*\Bigr]
  \end{split}
\end{equation*}
\begin{equation*}
  \begin{split}
    q_4 &= \frac{\hat{a}}{4}\Bigl[4e_f(1+11e_f^2)-2e_fe_p^2(34-33\sigma^2)-40e_fe_p\cos{\beta_*}\\
      &\quad -2e_fe_p^2(34+33\sigma^2)\cos 2\beta_*+e_p\sigma\bigl(4+132e_f^2-e_p^2(34\\
      &\quad -33\sigma^2)\bigr)\sin\beta_*-20e_p^2\sigma\sin 2\beta_*-e_p^3\sigma(34+11\sigma^2)\sin 3\beta_*\Bigr],
  \end{split}
\end{equation*}
\begin{equation*}
  \begin{split}
    q_5 &= 2\hat{a}\bigl[e_p^2(1-\sigma^2)-2e_f^2\bigr]+8K\kappa-\frac{1}{2}\hat{a}e_p\bigl[28e_f^2\\
      &\quad +e_p^2(7\sigma^2-6)-4\bigr]\cos\beta_*+2\hat{a}e_p^2(1+\sigma^2)\cos 2\beta_*\\
      &\quad +\hat{a}e_p^3\Bigl(1+\frac{7}{2}\sigma^2\Bigr)\cos 3\beta_*
      -2\hat{a}e_fe_p\sigma(4\sin\beta_*+7e_p\sin 2\beta_*),
  \end{split}
\end{equation*}
\begin{equation*}
  \begin{split}
    q_6 &= -\frac{\hat{a}}{4}\Bigl[4e_f(e_f^2-1)-2e_fe_p^2(2-3\sigma^2)-8e_fe_p\cos{\beta_*}\\
      &\quad -2e_fe_p^2(2+3\sigma^2)\cos 2\beta_*-e_p\sigma\bigl(4-12e_f^2+e_p^2(2\\
      &\quad -3\sigma^2)\bigr)\sin\beta_*-4e_p^2\sigma\sin 2\beta_*-e_p^3\sigma(2+\sigma^2)\sin 3\beta_*\Bigr].
  \end{split}
\end{equation*} 

\noindent
The coefficients of the polynomial $p(y)$, introduced
in~\eqref{eq:pq}, are given by:
\begin{equation*}
  \begin{split}
    p_0  &= \frac{\sigma}{4}\Bigl[-4e_f+2e_fe_p\cos\hat{\theta}_*
      -4e_f\cos 2\hat{\theta}_*+6e_fe_p\cos 3\hat{\theta}_*\\
      &\quad -(4-3e_f^2-e_p^2)\sin\hat{\theta}_*+4e_p\sin 2\hat{\theta}_*+3(e_f^2-e_p^2)\sin 3\hat{\theta}_*\Bigr], 
  \end{split}
\end{equation*}
\begin{equation*}
  \begin{split}
    p_1 &= \frac{1}{2}\Bigl[4e_p(1-\sigma^2)+\bigl(4-e_f^2+e_p^2(2\sigma^2-3)\bigr)\cos\hat{\theta}_*\\
      &\quad -4e_p(1+\sigma^2)\cos 2\hat{\theta}_*+3\bigl(e_p^2(1+2\sigma^2)-e_f^2\bigr)\cos 3\hat{\theta}_*\\
      &\quad -2e_fe_p(1-3\sigma^2)\sin\hat{\theta}_*-4e_f\sin 2\hat{\theta}_*\\
      & \quad\, +6e_fe_p(1+\sigma^2)\sin 3\hat{\theta}_*\Bigr],
  \end{split}
\end{equation*}
\begin{equation*}
  \begin{split}
    p_2 &= -\frac{\sigma}{4}\Bigl[4e_f+10e_fe_p\cos\hat{\theta}_*+4e_f\cos 2\hat{\theta}_*+30e_fe_p\cos 3\hat{\theta}_*\\
      &\quad +\bigl(4-3e_f^2+e_p^2(11-12\sigma^2)\bigr)\sin\hat{\theta}_*+20e_p\sin 2\hat{\theta}_*\\
      &\quad -3\bigl(e_f^2+e_p^2(11+4\sigma^2)\bigr)\sin 3\hat{\theta}_*\Bigr],
  \end{split}
\end{equation*}
\begin{equation*}
  \begin{split}
    p_3 &= 2\cos\hat{\theta}_*\Bigl[2+e_f^2+e_p^2(3+4\sigma^2)-3\bigl(e_f^2+e_p^2(1+4\sigma^2)\bigr)\cos 2\hat{\theta}_*\\
      &\quad -4e_f\sin \hat{\theta}_*\Bigr],
  \end{split}
\end{equation*}
\begin{equation*}
  \begin{split}
    p_4 &= \frac{\sigma}{4}\Bigl[4e_f-10e_fe_p\cos\hat{\theta}_*+4e_f\cos 2\hat{\theta}_*-30e_fe_p\cos 3\hat{\theta}_*\\
      &\quad +\bigl(4-3e_f^2+e_p^2(11-12\sigma^2)\bigr)\sin\hat{\theta}_*-20e_p\sin 2\hat{\theta}_*\\
      &\quad -3\bigl(e_f^2+e_p^2(11+4\sigma^2)\bigr)\sin 3\hat{\theta}_*\Bigr],
  \end{split}
\end{equation*}
\begin{equation*}
  \begin{split}
    p_5 &= \frac{1}{2}\Bigl[4e_p(\sigma^2-1)+\bigl(4-e_f^2+e_p^2(2\sigma^2-3)\bigr)\cos\hat{\theta}_*\\
      &\quad +4e_p(1+\sigma^2)\cos 2\hat{\theta}_*+3\bigl(e_p^2(1+2\sigma^2)-e_f^2\bigr)\cos 3\hat{\theta}_*\\
      &\quad +2e_fe_p(1-3\sigma^2)\sin\hat{\theta}_*-4e_f\sin 2\hat{\theta}_*\\
      &\quad\, -6e_fe_p(1+\sigma^2)\sin 3\hat{\theta}_*\Bigr],
  \end{split}
\end{equation*}
\begin{equation*}
  \begin{split}
    p_6  &= \frac{\sigma}{4}\Bigl[4e_f+2e_fe_p\cos\hat{\theta}_*+4e_f\cos 2\hat{\theta}_*+6e_fe_p\cos 3\hat{\theta}_*\\
      &\quad +(4-3e_f^2-e_p^2)\sin\hat{\theta}_*+4e_p\sin 2\hat{\theta}_*-3(e_f^2-e_p^2)\sin 3\hat{\theta}_*\Bigr]. 
  \end{split}
\end{equation*}

\section{Values of the angular and radial buffers}
\label{ap:buf}

The values of the angular buffer depend on the mutual inclination of
the pair and the orbit inclination $\hat{i}$ of each object.
Tables~\ref{tab:theta_bf_0} and~\ref{tab:theta_bf_f} report the values
at the initial time $t_0$ and final time $t_{\rm f}$, respectively.

\begin{table*}
\centering
\begin{turn}{90}
\begin{adjustbox}{width=\columnwidth,center}
\begin{tabular}{lrrrrrrrrrrrrrrr}
\hline
& \multicolumn{15}{c}{Orbit inclination (deg)} \\
\cline{2-16}
\makecell[l]{Mutual\\ inclination (deg)} & {\scriptsize $[0,10]$} & {\scriptsize $(10,20]$} & {\scriptsize$(20,30]$}
  & {\scriptsize$(30,40]$} & {\scriptsize$(40,50]$} & {\scriptsize$(50,60]$} & {\scriptsize$(60,70]$} & {\scriptsize$(70,80]$}
  & {\scriptsize$(80,90]$} & {\scriptsize$(90,100]$} & {\scriptsize$(100,110]$} & {\scriptsize$(110,120]$} & {\scriptsize$(120,130]$}
  & {\scriptsize$(130,140]$} & {\scriptsize$(140,150]$} \\
\hline
{\scriptsize $[10,15]$} & 0.114 & 0.181 & 0.303 & 0.226 & 0.237 & 0.241 & 0.201 & 0.156 & 0.102 & 0.104 & 0.133 & 0.138 & 0.140 & 0 & 0.367 \\
{\scriptsize $(15,20]$} & 0.175 & 0.158 & 0.323 & 0.176 & 0.221 & 0.158 & 0.140 & 0.107 & 0.083 & 0.092 & 0.093 & 0.097 & 0.125 & 0 & 0.130 \\
{\scriptsize $(20,25]$} & 0.164 & 0.113 & 0.142 & 0.132 & 0.119 & 0.124 & 0.091 & 0.081 & 0.064 & 0.075 & 0.075 & 0.075 & 0.087 & 0 & 0.095 \\
{\scriptsize $(25,30]$} & 0.107 & 0.086 & 0.158 & 0.111 & 0.096 & 0.099 & 0.082 & 0.064 & 0.053 & 0.059 & 0.069 & 0.060 & 0.067 & 0 & 0.097 \\
{\scriptsize $(30,35]$} & 0.087 & 0.084 & 0.101 & 0.138 & 0.079 & 0.084 & 0.070 & 0.054 & 0.050 & 0.057 & 0.051 & 0.052 & 0.062 & 0 & 0.084 \\
{\scriptsize $(35,40]$} & 0.102 & 0.080 & 0.090 & 0.093 & 0.072 & 0.075 & 0.059 & 0.050 & 0.045 & 0.042 & 0.045 & 0.041 & 0.045 & 0 & 0.067 \\
{\scriptsize $(40,45]$} & 0.066 & 0.071 & 0.084 & 0.067 & 0.057 & 0.064 & 0.053 & 0.043 & 0.035 & 0.039 & 0.040 & 0.036 & 0.040 & 0 & 0.065 \\
{\scriptsize $(45,50]$} & 0.071 & 0.065 & 0.074 & 0.064 & 0.054 & 0.059 & 0.045 & 0.038 & 0.033 & 0.035 & 0.034 & 0.034 & 0.034 & 0 & 0.052 \\
{\scriptsize $(50,55]$} & 0.061 & 0.063 & 0.066 & 0.062 & 0.046 & 0.054 & 0.042 & 0.035 & 0.030 & 0.030 & 0.031 & 0.027 & 0.030 & 0 & 0.049 \\
{\scriptsize $(55,60]$} & 0.065 & 0.059 & 0.084 & 0.057 & 0.044 & 0.049 & 0.038 & 0.034 & 0.026 & 0.028 & 0.029 & 0.025 & 0.031 & 0 & 0.047 \\
{\scriptsize $(60,65]$} & 0.059 & 0.056 & 0.074 & 0.052 & 0.049 & 0.047 & 0.032 & 0.032 & 0.026 & 0.025 & 0.027 & 0.020 & 0.025 & 0 & 0.043 \\
{\scriptsize $(65,70]$} & 0.057 & 0.055 & 0.076 & 0.048 & 0.050 & 0.044 & 0.033 & 0.029 & 0.026 & 0.026 & 0.026 & 0.023 & 0.026 & 0 & 0.044 \\
{\scriptsize $(70,75]$} & 0.054 & 0.053 & 0.073 & 0.046 & 0.051 & 0.042 & 0.029 & 0.025 & 0.026 & 0.025 & 0.026 & 0.023 & 0.026 & 0 & 0.043 \\
{\scriptsize $(75,80]$} & 0.049 & 0.052 & 0.053 & 0.046 & 0.043 & 0.041 & 0.030 & 0.024 & 0.026 & 0.025 & 0.022 & 0.020 & 0.027 & 0 & 0.043 \\
{\scriptsize $(80,85]$} & 0.046 & 0.053 & 0.085 & 0.049 & 0.041 & 0.053 & 0.029 & 0.023 & 0.024 & 0.023 & 0.024 & 0.027 & 0.029 & 0 & 0.046 \\
{\scriptsize $(85,90]$} & 0.044 & 0.049 & 0.056 & 0.049 & 0.040 & 0.041 & 0.031 & 0.024 & 0.022 & 0.024 & 0.025 & 0.030 & 0.034 & 0 & 0.045 \\
{\scriptsize $(90,95]$} & 0.040 & 0.050 & 0.058 & 0.049 & 0.043 & 0.042 & 0.032 & 0.026 & 0.023 & 0.024 & 0.028 & 0.028 & 0.037 & 0 & 0.045 \\
{\scriptsize $(95,100]$} & 0.037 & 0.050 & 0.053 & 0.049 & 0.045 & 0.044 & 0.038 & 0.027 & 0.024 & 0.026 & 0.028 & 0.027 & 0.039 & 0 & 0.046 \\
{\scriptsize $(100,105]$} & 0 & 0.049 & 0.053 & 0.048 & 0.043 & 0.046 & 0.035 & 0.030 & 0.026 & 0.028 & 0.030 & 0.032 & 0.036 & 0 & 0.042 \\
{\scriptsize $(105,110]$} & 0 & 0.047 & 0.052 & 0.049 & 0.044 & 0.047 & 0.040 & 0.030 & 0.027 & 0.030 & 0.032 & 0.034 & 0.037 & 0 & 0.042 \\
{\scriptsize $(110,115]$} & 0 & 0 & 0.050 & 0.047 & 0.045 & 0.046 & 0.041 & 0.032 & 0.028 & 0.032 & 0.033 & 0.032 & 0.034 & 0 & 0.040 \\
{\scriptsize $(115,120]$} & 0 & 0 & 0.049 & 0.046 & 0.045 & 0.047 & 0.043 & 0.033 & 0.029 & 0.034 & 0.037 & 0.039 & 0.033 & 0 & 0.039 \\
{\scriptsize $(120,125]$} & 0 & 0 & 0 & 0.045 & 0.042 & 0.049 & 0.044 & 0.037 & 0.034 & 0.038 & 0.038 & 0.039 & 0.037 & 0 & 0.037 \\
{\scriptsize $(125,130]$} & 0 & 0 & 0 & 0.044 & 0.042 & 0.050 & 0.052 & 0.039 & 0.037 & 0.040 & 0.042 & 0.039 & 0.034 & 0 & 0 \\
{\scriptsize $(130,135]$} & 0 & 0 & 0 & 0 & 0.040 & 0.050 & 0.050 & 0.044 & 0.039 & 0.043 & 0.043 & 0.037 & 0.042 & 0 & 0 \\
{\scriptsize $(135,140]$} & 0 & 0 & 0 & 0 & 0.036 & 0.049 & 0.050 & 0.048 & 0.039 & 0.046 & 0.043 & 0.037 & 0.043 & 0 & 0 \\
{\scriptsize $(140,145]$} & 0 & 0 & 0 & 0 & 0 & 0.048 & 0.050 & 0.044 & 0.043 & 0.046 & 0.043 & 0.037 & 0.039 & 0 & 0 \\
{\scriptsize $(145,150]$} & 0 & 0 & 0 & 0 & 0 & 0.039 & 0.050 & 0.045 & 0.045 & 0.045 & 0.044 & 0.037 & 0.030 & 0 & 0 \\
{\scriptsize $(150,155]$} & 0 & 0 & 0 & 0 & 0 & 0 & 0.050 & 0.047 & 0.045 & 0.045 & 0.042 & 0.036 & 0 & 0 & 0 \\
{\scriptsize $(155,160]$} & 0 & 0 & 0 & 0 & 0 & 0 & 0.047 & 0.046 & 0.057 & 0.053 & 0.051 & 0 & 0 & 0 & 0 \\
{\scriptsize $(160,165]$} & 0 & 0 & 0 & 0 & 0 & 0 & 0 & 0.049 & 0.060 & 0.062 & 0.047 & 0 & 0 & 0 & 0 \\
{\scriptsize $(165,170]$} & 0 & 0 & 0 & 0 & 0 & 0 & 0 & 0.042 & 0.052 & 0.062 & 0.044 & 0 & 0 & 0 & 0 \\
\hline
\end{tabular}
\end{adjustbox}
\end{turn}
\caption{Values of the angular buffer (in degrees) at $t_0$.}
\label{tab:theta_bf_0}
\end{table*}

\begin{table*}
\centering
\begin{turn}{90}
\begin{adjustbox}{width=\columnwidth,center}
\begin{tabular}{lrrrrrrrrrrrrrrr}
\hline
& \multicolumn{15}{c}{Orbit inclination (deg)} \\
\cline{2-16}
\makecell[l]{Mutual\\ inclination (deg)} & {\scriptsize$[0,10)$} & {\scriptsize $(10,20]$} & {\scriptsize$(20,30]$}
  & {\scriptsize$(30,40]$} & {\scriptsize$(40,50]$} & {\scriptsize$(50,60]$} & {\scriptsize$(60,70]$} & {\scriptsize$(70,80]$}
  & {\scriptsize$(80,90]$} & {\scriptsize$(90,100]$} & {\scriptsize$(100,110]$} & {\scriptsize$(110,120]$} & {\scriptsize$(120,130]$}
  & {\scriptsize$(130,140]$} & {\scriptsize$(140,150]$} \\
\hline
{\scriptsize $[10,15]$} & 0.253 & 0.305 & 0.387 & 0.363 & 0.339 & 0.294 & 0.332 & 0.331 & 0.249 & 0.316 & 0.309 & 0.285 & 0.232 & 0 & 0.240 \\
{\scriptsize $(15,20]$} & 0.533 & 0.314 & 0.323 & 0.323 & 0.216 & 0.235 & 0.270 & 0.280 & 0.165 & 0.217 & 0.246 & 0.180 & 0.312 & 0 & 0.153 \\
{\scriptsize $(20,25]$} & 0.530 & 0.298 & 0.311 & 0.267 & 0.177 & 0.195 & 0.203 & 0.124 & 0.131 & 0.183 & 0.162 & 0.144 & 0.150 & 0 & 0.123 \\
{\scriptsize $(25,30]$} & 0.461 & 0.298 & 0.282 & 0.223 & 0.139 & 0.189 & 0.151 & 0.105 & 0.124 & 0.138 & 0.154 & 0.096 & 0.144 & 0 & 0.121 \\
{\scriptsize $(30,35]$} & 0.388 & 0.292 & 0.261 & 0.192 & 0.136 & 0.165 & 0.119 & 0.092 & 0.082 & 0.133 & 0.119 & 0.071 & 0.100 & 0 & 0.086 \\
{\scriptsize $(35,40]$} & 0.385 & 0.303 & 0.219 & 0.179 & 0.124 & 0.127 & 0.104 & 0.096 & 0.078 & 0.094 & 0.082 & 0.061 & 0.074 & 0 & 0.112 \\
{\scriptsize $(40,45]$} & 0.327 & 0.279 & 0.199 & 0.164 & 0.107 & 0.108 & 0.089 & 0.089 & 0.066 & 0.075 & 0.069 & 0.055 & 0.067 & 0 & 0.093 \\
{\scriptsize $(45,50]$} & 0.317 & 0.275 & 0.206 & 0.149 & 0.091 & 0.087 & 0.083 & 0.085 & 0.065 & 0.058 & 0.056 & 0.041 & 0.059 & 0 & 0.094 \\
{\scriptsize $(50,55]$} & 0.339 & 0.256 & 0.182 & 0.125 & 0.084 & 0.075 & 0.069 & 0.074 & 0.066 & 0.052 & 0.052 & 0.037 & 0.054 & 0 & 0.085 \\
{\scriptsize $(55,60]$} & 0.331 & 0.234 & 0.168 & 0.107 & 0.069 & 0.069 & 0.063 & 0.062 & 0.063 & 0.049 & 0.047 & 0.031 & 0.057 & 0 & 0.078 \\
{\scriptsize $(60,65]$} & 0.320 & 0.221 & 0.165 & 0.108 & 0.068 & 0.076 & 0.052 & 0.052 & 0.058 & 0.049 & 0.046 & 0.036 & 0.052 & 0 & 0.078 \\
{\scriptsize $(65,70]$} & 0.319 & 0.218 & 0.146 & 0.103 & 0.078 & 0.067 & 0.048 & 0.041 & 0.055 & 0.048 & 0.036 & 0.034 & 0.058 & 0 & 0.068 \\
{\scriptsize $(70,75]$} & 0.311 & 0.209 & 0.142 & 0.097 & 0.066 & 0.068 & 0.056 & 0.038 & 0.052 & 0.045 & 0.038 & 0.045 & 0.047 & 0 & 0.066 \\
{\scriptsize $(75,80]$} & 0.308 & 0.202 & 0.143 & 0.099 & 0.064 & 0.071 & 0.058 & 0.048 & 0.034 & 0.036 & 0.043 & 0.042 & 0.053 & 0 & 0.059 \\
{\scriptsize $(80,85]$} & 0.298 & 0.206 & 0.144 & 0.107 & 0.063 & 0.073 & 0.064 & 0.051 & 0.040 & 0.038 & 0.045 & 0.044 & 0.053 & 0 & 0.058 \\
{\scriptsize $(85,90]$} & 0.285 & 0.202 & 0.146 & 0.099 & 0.068 & 0.091 & 0.065 & 0.056 & 0.046 & 0.042 & 0.049 & 0.047 & 0.077 & 0 & 0.064 \\
{\scriptsize $(90,95]$} & 0.274 & 0.199 & 0.138 & 0.107 & 0.077 & 0.099 & 0.065 & 0.060 & 0.050 & 0.057 & 0.055 & 0.057 & 0.090 & 0 & 0.064 \\
{\scriptsize $(95,100]$} & 0.231 & 0.197 & 0.139 & 0.104 & 0.074 & 0.088 & 0.062 & 0.060 & 0.053 & 0.056 & 0.064 & 0.062 & 0.081 & 0 & 0.068 \\
{\scriptsize $(100,105]$} & 0 & 0.188 & 0.133 & 0.102 & 0.078 & 0.098 & 0.078 & 0.062 & 0.055 & 0.063 & 0.068 & 0.065 & 0.091 & 0 & 0.069 \\
{\scriptsize $(105,110]$} & 0 & 0.112 & 0.129 & 0.100 & 0.085 & 0.097 & 0.074 & 0.058 & 0.057 & 0.072 & 0.077 & 0.066 & 0.056 & 0 & 0.075 \\
{\scriptsize $(110,115]$} & 0 & 0 & 0.120 & 0.108 & 0.087 & 0.110 & 0.090 & 0.090 & 0.056 & 0.084 & 0.081 & 0.074 & 0.096 & 0 & 0.076 \\
{\scriptsize $(115,120]$} & 0 & 0 & 0.101 & 0.092 & 0.093 & 0.110 & 0.096 & 0.091 & 0.062 & 0.085 & 0.090 & 0.078 & 0.100 & 0 & 0.073 \\
{\scriptsize $(120,125]$} & 0 & 0 & 0 & 0.078 & 0.095 & 0.114 & 0.103 & 0.081 & 0.060 & 0.090 & 0.093 & 0.084 & 0.093 & 0 & 0.055 \\
{\scriptsize $(125,130]$} & 0 & 0 & 0 & 0.038 & 0.092 & 0.123 & 0.112 & 0.088 & 0.083 & 0.098 & 0.099 & 0.091 & 0.094 & 0 & 0 \\
{\scriptsize $(130,135]$} & 0 & 0 & 0 & 0 & 0.076 & 0.116 & 0.121 & 0.100 & 0.071 & 0.106 & 0.104 & 0.094 & 0.064 & 0 & 0 \\
{\scriptsize $(135,140]$} & 0 & 0 & 0 & 0 & 0.072 & 0.095 & 0.126 & 0.111 & 0.085 & 0.114 & 0.107 & 0.093 & 0.091 & 0 & 0 \\
{\scriptsize $(140,145]$} & 0 & 0 & 0 & 0 & 0 & 0.087 & 0.124 & 0.121 & 0.091 & 0.115 & 0.118 & 0.095 & 0.051 & 0 & 0 \\
{\scriptsize $(145,150]$} & 0 & 0 & 0 & 0 & 0 & 0.064 & 0.123 & 0.136 & 0.107 & 0.132 & 0.126 & 0.090 & 0.031 & 0 & 0 \\
{\scriptsize $(150,155]$} & 0 & 0 & 0 & 0 & 0 & 0 & 0.114 & 0.140 & 0.142 & 0.141 & 0.144 & 0.083 & 0 & 0 & 0 \\
{\scriptsize $(155,160]$} & 0 & 0 & 0 & 0 & 0 & 0 & 0.100 & 0.139 & 0.157 & 0.158 & 0.146 & 0 & 0 & 0 & 0 \\
{\scriptsize $(160,165]$} & 0 & 0 & 0 & 0 & 0 & 0 & 0 & 0.123 & 0.167 & 0.167 & 0.137 & 0 & 0 & 0 & 0 \\
{\scriptsize $(165,170]$} & 0 & 0 & 0 & 0 & 0 & 0 & 0 & 0.122 & 0.173 & 0.174 & 0.129 & 0 & 0 & 0 & 0\\
\hline
\end{tabular}
\end{adjustbox}
\end{turn}
\caption{Values of the angular buffer (in degrees) at $t_{\rm f}$.}
\label{tab:theta_bf_f}
\end{table*}

The values of the radial buffer depend on the semi-major axis
$\hat{a}$, eccentricity $\hat{e}$, and the precession period $T_C$ of
the apsidal line. Tables~\ref{tab:R_bf_high} and~\ref{tab:R_bf_low}
report the values for $\hat{a}>10,000$~km and $\hat{a}\le10,000$~km,
respectively.

\begin{table}
\centering
\begin{tabular}{lrr}
\hline
& \multicolumn{2}{c}{Apsidal line period (days)} \\
\cline{2-3}
Eccentricity & $(0,350]$ & $>350$ \\
\hline
$[0,0.01]$ & 0 & 2.154 \\
$(0.01,0.02]$ & 0 & 2.239 \\
$(0.02,0.03]$ & 0 & 2.445 \\
$(0.03,0.04]$ & 0 & 2.449 \\
$(0.04,0.05]$ & 0 & 2.412 \\
$(0.05,0.06]$ & 0 & 2.403 \\
$(0.06,0.07]$ & 0.497 & 0.510 \\
$(0.07,0.08]$ & 0.599 & 4.876 \\
$(0.08,0.09]$ & 0 & 0 \\
$(0.09,0.1]$ & 0 & 0 \\
\hline
\end{tabular}
\caption{Values of the radial buffer (in km) for
  $\hat{a}>10,000$~km. All orbits in the dataset satisfying this
  condition on $\hat{a}$ have $T_C>300$~days.}
\label{tab:R_bf_high}
\end{table}

\begin{table}
\begin{adjustbox}{width=\columnwidth,center}
\begin{tabular}{lrrrrrrrr}
\hline
& \multicolumn{8}{c}{Apsidal line period (days)} \\
\cline{2-9}
Eccentricity & $(0,50]$ & $(50,100]$ & $(100,150]$ & $(150,200]$ & $(200,250]$ & $(250,300]$ & $(300,350]$ & $>350$ \\
\hline
$[0,0.01]$ & 1.093 & 1.078 & 0.752 & 0.719 & 0.603 & 0.580 & 0.679 & 0.752 \\
$(0.01,0.02]$ & 0.925 & 0.635 & 0.900 & 0.711 & 0.625 & 0.588 & 0.713 & 0.767 \\
$(0.02,0.03]$ & 1.089 & 0.733 & 0.848 & 0.667 & 0.516 & 0.635 & 0.367 & 0.855 \\
$(0.03,0.04]$ & 2.241 & 1.017 & 0.836 & 0.858 & 0.637 & 0.384 &     0 & 1.073 \\
$(0.04,0.05]$ & 2.436 & 1.113 & 0.878 & 0.670 & 0.936 & 0.639 &     0 & 1.161 \\
$(0.05,0.06]$ & 1.970 & 1.700 & 0.948 & 0.708 & 1.178 &     0 &     0 & 1.123 \\
$(0.06,0.07]$ & 2.805 & 1.969 & 1.066 & 0.964 & 1.160 & 0.577 & 1.147 & 1.362 \\
$(0.07,0.08]$ & 2.641 & 2.179 & 0.874 & 1.028 & 0.801 & 0.971 & 0.605 & 1.558 \\
$(0.08,0.09]$ & 3.730 & 2.535 & 1.067 & 0.808 & 0.752 & 1.369 & 0.680 & 1.636 \\
$(0.09,0.1]$  & 2.794 & 2.574 & 1.787 & 1.608 & 1.503 &     0 &     0 & 1.780 \\
\hline
\end{tabular}
\end{adjustbox}
\caption{Values of the radial buffer (in km) for
  $\hat{a}\le10,000$~km.}
\label{tab:R_bf_low}
\end{table}

\end{document}